\documentclass[11pt,a4paper]{article}
\pdfoutput=1
\usepackage{jheppub}

\usepackage{color}
\usepackage{amsmath}
\usepackage{pifont}   
\usepackage{verbatim}       
\usepackage{acronym} 
    
\usepackage{amsfonts}
\usepackage{amssymb}
\usepackage{mathrsfs} 
\usepackage{graphicx}
\usepackage{multirow} 
\usepackage{slashed} 
\usepackage{array}

\usepackage{graphicx}
\usepackage{epsfig}
\usepackage{lscape}
\usepackage{rotating}
\usepackage{epsf}
\usepackage{bm}
\usepackage{booktabs}
\usepackage{array}
\usepackage{caption}
\usepackage{subcaption}
\usepackage[utf8]{inputenc}
\usepackage{float}

\usepackage{multirow}
\usepackage{array,booktabs,colortbl,multirow}
\usepackage{colortbl,xcolor,colordvi,color}
\usepackage{rotating}
\allowdisplaybreaks

\newcommand{\cmrule}{\midrule[0.25mm]}

\title{Measuring properties of a dark photon from semi-invisible decay of the Higgs boson}

\author[a]{Hugues Beauchesne}
\author[b,a]{and Cheng-Wei Chiang}

\affiliation[a]{Physics Division, National Center for Theoretical Sciences,\\ Taipei 10617, Taiwan}
\affiliation[b]{Department of Physics and Center for Theoretical Physics, National Taiwan University, \\ Taipei 10617, Taiwan}

\emailAdd{beauchesneh@phys.ncts.ntu.edu.tw, chengwei@phys.ntu.edu.tw}

\abstract{Considerable efforts have been dedicated to discovering a dark photon via the decay of the Higgs boson to a photon and an invisible particle. A subject that is still mostly unexplored is which properties of the dark photon could be measured at the LHC if an excess were to be found in this channel and whether we could determine if this signal is indeed that of a dark photon. In this paper, we seek to address some of these questions for two Higgs production channels: gluon-fusion and $Z$-associated. First, prospects are presented for the upper limit on the mass of a massless dark photon and for the uncertainty on the mass of a massive dark photon. Second, we study the feasibility of distinguishing this signal from that of the Higgs decaying to a gravitino and a neutralino that decays to a photon and another gravitino. Finally, the complementary possibility of observing the decay of the Higgs to a dark photon and a $Z$ boson is studied.}

\begin{document}

\maketitle

\section{Introduction}\label{Sec:Intro}

A plethora of experimental evidence leaves little doubt about the existence of dark matter \cite{Planck:2018vyg}, but its exact nature remains unknown. Since dark matter was first proposed, most experimental efforts have been directed at Weakly Interacting Massive Particles (WIMPs), particles with masses around the electroweak scale and couplings of similar magnitude to the electroweak ones. Due to the absence of unambiguous WIMP signals, there has been considerable efforts in recent years to search for alternative dark matter candidates. One of the most studied amongst these is the dark photon \cite{Holdom:1985ag}, a new Abelian gauge boson that mixes with the photon via kinetic mixing.  The dark photon may also play the role of a messenger between the visible and dark sectors~\cite{Chiang:2020hgb}.

One possible discovery channel that has received much attention is the decay of the Higgs boson to a photon and a dark photon. Theoretical studies include Refs.~\cite{Gopalakrishna:2008dv, Davoudiasl:2012ag, Curtin:2013fra, Curtin:2014cca, Gabrielli:2014oya, Biswas:2015sha, Biswas:2016jsh} and experimental studies include Refs.~\cite{ATLAS:2015hpr, ATLAS:2018coo, CMS-PAS-HIG-19-007, ATLAS:2019tkk, CMS:2019ajt, CMS:2019buh, CMS:2020krr, ATLAS:2021pdg}. This channel is especially favoured, as a branching ratio of the Higgs to a photon and a dark photon of a few percents is compatible with other constraints and within the reach of current collider experiments \cite{Gabrielli:2014oya}.

An aspect that is still unclear is, if we indeed observe the decay of the Higgs boson to a photon and a dark photon, how precisely could we measure the properties of the dark photon and could we even distinguish it from alternative hypotheses.

In this paper, we seek to address some of these questions. More precisely, we will consider what properties of the dark photon could be inferred from the semi-invisible decay of the Higgs boson to a photon and a stable dark photon if an excess is found at the LHC.\footnote{For searches at lepton colliders, please see, for example, Ref.~\cite{He:2017zzr}.}  We will consider two Higgs production channels: gluon-fusion and $Z$-associated (also known as Higgsstrahlung). These two benchmark channels are chosen for their contrasting natures and will illustrate different scenarios. The gluon-fusion production has a larger background, but a discovery would imply a large amount of signals available to constrain the properties of the dark photon. The $Z$-associated channel has very little background, but a discovery would have fewer signal events to work with.

We will concentrate on three main questions. How well could we measure the mass of the dark photon? How can we tell the excess apart from the well-motivated case of the Higgs decaying to a gravitino and a neutralino that in turn decays to a photon and another gravitino? Is the observation of this excess compatible with the lack of observation of the Higgs decaying to a dark photon and a $Z$ boson?

We have found the following results. For a massless dark photon, the LHC could impose an upper limit on its mass of a few GeV in the best case scenario. For a massive one, its mass could potentially be measured up to sub-GeV precision, depending on how heavy it is. The alternative hypothesis of the Higgs decaying to a light gravitino and a neutralino decaying to a photon and another gravitino could potentially be excluded at close to 95\% confidence level (CL). In general, the gluon-fusion channel leads to better results than the $Z$-associated production. Finally, the decay of the Higgs to a dark photon and a $Z$ could go undetected without being in conflict with the observation of an excess in the decay to a dark photon and a photon.

The paper is organized as follows. First, we elaborate on the technical details of our simulations and the relevant backgrounds in Section~\ref{Sec:BackgroundSimulation}. In Section~\ref{Sec:MassDetermination}, prospects for the upper limit or uncertainty on the mass of a dark photon are presented. The ability to exclude the neutralino/gravitino alternative hypothesis is analyzed afterward in Section~\ref{Sec:Sec:N1N2}. Section~\ref{Sec:HiggstoZgammaD} is devoted to the discussion of the decay of the Higgs to a dark photon and $Z$. Finally, we present some concluding remarks in Section~\ref{Sec:Conclusion}.  Appendix~\ref{Sec:DecayWidths} collects the formulas of the masses and mixing of the neutral gauge bosons and the decays of $h \to A A'$ and $Z A'$.

\section{Background and simulation details}\label{Sec:BackgroundSimulation}

We begin by describing the event generation and discussing the relevant backgrounds for each Higgs production channel considered in this work.

\subsection{General comments}\label{sSec:GeneralComments}

All events are generated using \texttt{MadGraph} 2 \cite{Alwall:2014hca} and an implementation of the relevant models in \texttt{FeynRules} \cite{Alloul:2013bka}. Parton showering and hadronization is handled through \texttt{PYTHIA}~8 \cite{Sjostrand:2007gs}. Detector simulation is done with \texttt{Delphes} 3 \cite{deFavereau:2013fsa} using the CMS settings. The only exceptions to this are the photon identification efficiency, which is set to a value presented below, and the photon isolation requirements, which are set to emulate those of Ref.~\cite{ATL-PHYS-PUB-2016-026}. Unless stated otherwise, the cross sections are computed using the next-to-leading (NLO) functionality of \texttt{MadGraph}. All results of this paper are presented for a center-of-mass energy of 14~TeV.

Some of the most important backgrounds will prove to be jets and electrons mistagged as photons. For electrons, the mistagging rate is set to 2\%, which is a typical value for the tight identification and isolation requirements generally used in searches for a Higgs boson decaying to a photon and an invisible particle (see, for example, Refs.~\cite{CMS:2015ifd, CMS:2018ffd, CMS:2019ajt, CMS:2020krr, ATLAS-CONF-2021-004}). Since electrons mistagged as photons will never be a particularly strong background, the exact value of this mistagging rate is not expected to affect the final results much. The mistagging rate of jets as photons is taken from Ref.~\cite{ATL-PHYS-PUB-2016-026}. In practice, we use the parametrization of Ref.~\cite{Goncalves:2018qas} given by
\begin{equation}\label{eq:J->gammaMistagRate}
\epsilon_{j\to A} = \left\{ \begin{matrix} 
\displaystyle
5.3  \times 10^{-4} \exp\left[-6.5\left(\frac{p_T}{60.4\;\text{GeV}} - 1\right)^2\right], 
& \qquad \quad &  p_T < 65\;\text{GeV}, \\ 
\displaystyle
0.88 \times 10^{-4} \left[\exp\left(-\frac{p_T}{943\;\text{GeV}}\right) +\frac{248\;\text{GeV}}{p_T}\right], 
& \qquad \quad &  p_T > 65\;\text{GeV}.
\end{matrix} \right. 
\end{equation}
The corresponding photon identification efficiency is also taken from Ref.~\cite{ATL-PHYS-PUB-2016-026} using the parametrization of Ref.~\cite{Goncalves:2018qas} given by
\begin{equation}\label{eq:gamma->gammaRate}
  \epsilon_{A\to A} = 0.863 - 1.07\exp\left(-\frac{p_T}{34.8\;\text{GeV}}\right),
\end{equation}
whose range of validity is respected for all photons considered in this work.

\subsection{Backgrounds for gluon-fusion production}\label{sSec:GluonFusionBackgrounds}

The most important backgrounds for the gluon-fusion channel are presented in Table~\ref{table:GluonFusionBackgrounds}, including the number of simulated events for each of them.

\begin{table}[t]
	\begin{subfigure}{.5\linewidth}
      \caption{Gluon-fusion}
	  {\footnotesize
		\setlength\tabcolsep{5pt}
		\begin{center}
	      \begin{tabular}{lc}
	        \toprule
	        Background             & \#Events         \\
		    \cmrule
		    jets $+\; A$        & $10^7$           \\
            jet $\to A$            & $10^7$           \\
            $e \to A$              & $5\times 10^6$   \\
            $Z A$                  & $2.5\times 10^6$ \\
            $W A$                  & $2.5\times 10^6$ \\
            $W \to \mu (\tau) \nu$ & $2.5\times 10^6$ \\
            $AA$                   & $2.5\times 10^6$ \\
            $W \to e  \nu$         & $5\times 10^6$   \\
            \bottomrule
	      \end{tabular}
	    \end{center}
	  }
      \label{table:GluonFusionBackgrounds}
	\end{subfigure}
	\begin{subfigure}{.5\linewidth}
	  \caption{$Z$-associated}
	  {\footnotesize
		\setlength\tabcolsep{5pt}
		\begin{center}
	      \begin{tabular}{lc}
	        \toprule
	        Background             & \#Events         \\
		    \cmrule
		    $WZ\;(e \to A)$        & $10^7$           \\
		    $WZ$                   & $10^7$           \\
            $ZZ$                   & $5\times 10^6$   \\
            $WW$                   & $5\times 10^6$   \\
            $t\bar{t}$             & $5\times 10^6$   \\
            $ZA$                   & $5\times 10^6$   \\
            \bottomrule
	      \end{tabular}
	    \end{center}
	  }
      \label{table:Zassociated}
	\end{subfigure}
	\caption{List of dominant backgrounds and corresponding number of simulated events for (a) the gluon-fusion and (b) the $Z$-associated channels.} 
	\label{table:Backgrounds}
\end{table}

Several comments are in order. The jets $+\; A$ background is notoriously difficult to simulate and is not expected to give accurate predictions. To account for this, we follow Refs.~\cite{CMS:2015ifd, Biswas:2016jsh} and use multiplicative correction factors of 1.7 and 1.1 for 0 or 1 jet, respectively. For the jet $\to A$ mistagging background, the events simulated in \texttt{MadGraph} are $p p \to j j$. Considering the small mistagging probability and to avoid having to generate a prohibitively large number of events, a reweighting of the events is performed as follows. If an event contains jets that could potentially be mistagged as valid photons, one of the candidate jets is selected randomly with probability proportional to Eq.~\eqref{eq:J->gammaMistagRate} and is treated as a photon. The event is then given a weight corresponding to the probability of any candidate jet being mistagged.\footnote{This neglects the possibility of two jets being mistagged, but this is negligibly unlikely.} The same procedure is applied for the electron mistagging. A k-factor of 1.2 is taken for the $p p \to j j$ cross section \cite{Bellm:2019yyh}. For the $e \to A$ mistagging background, the events generated in \texttt{MadGraph} are the resonant production of an electronically decaying $W$ boson. This background contains specifically the events that pass selection cuts because of electron misidentification. The events that pass the selection cuts without mistagging are instead accounted in the $W \to e  \nu$ background. For the $Z A$ background, the $Z$ boson is always decayed to neutrinos. For the $W A$ background, the $W$ boson is always decayed leptonically (including to $\tau$).

\begin{figure}[t!]
\begin{center}
 \begin{subfigure}{0.49\textwidth}
    \centering
    \caption{Gluon-fusion}
    \includegraphics[width=1\textwidth]{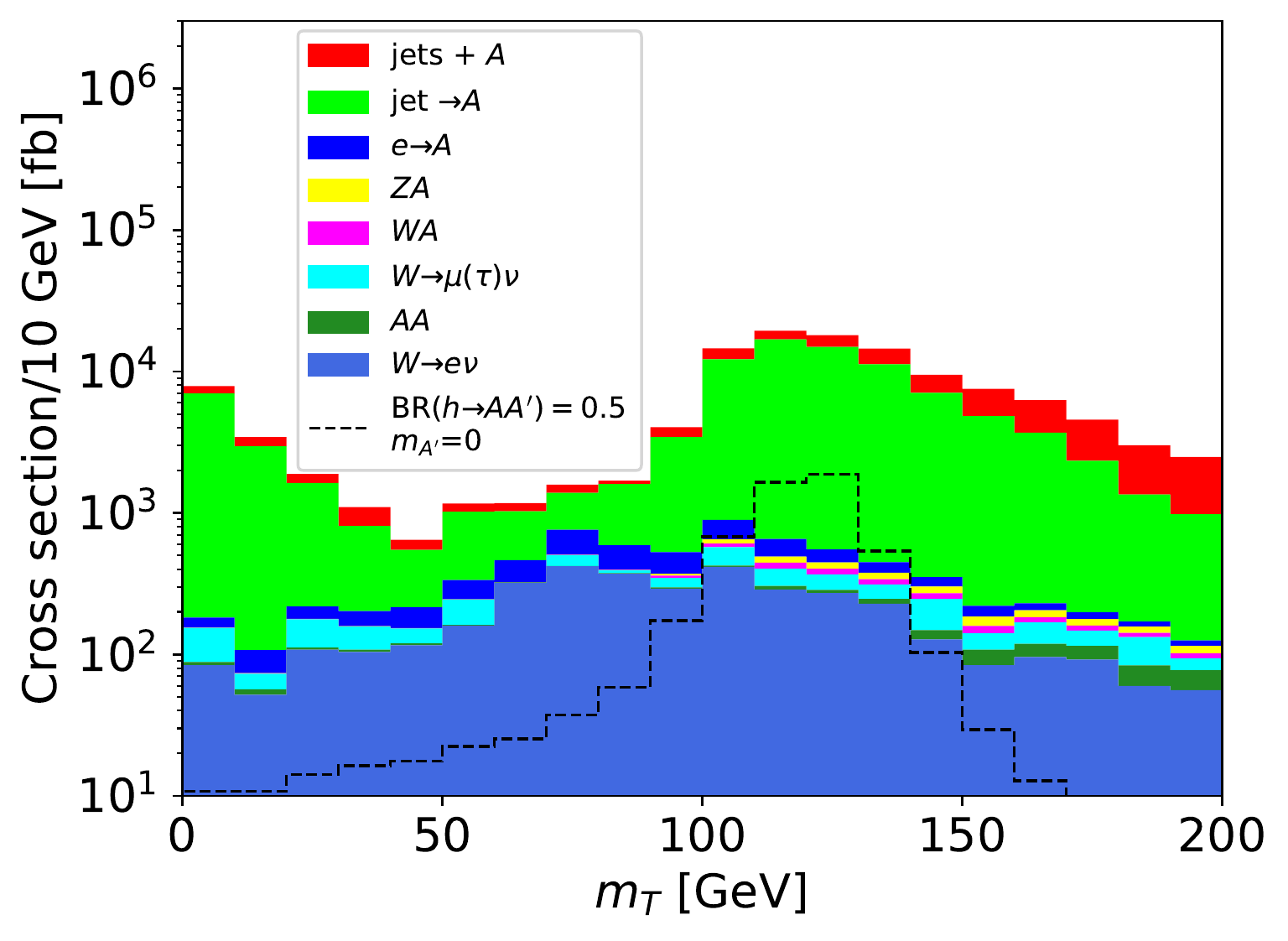}
    \label{fig:BackGroundsGF}
  \end{subfigure}
 \begin{subfigure}{0.49\textwidth}
    \centering
    \caption{$Z$-associated}
    \includegraphics[width=1\textwidth]{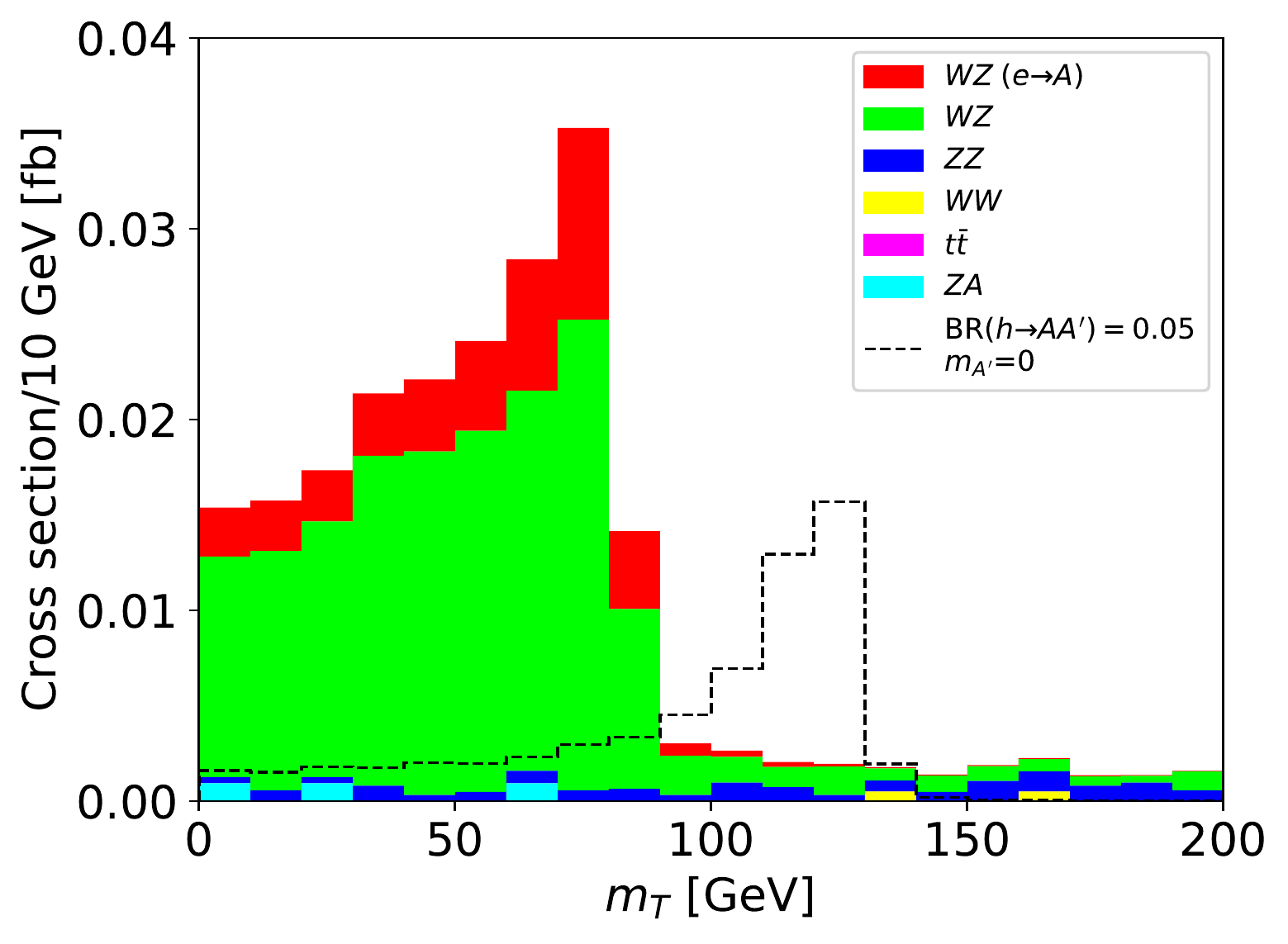}
    \label{fig:BackGroundsZH}
  \end{subfigure}
\caption{The $m_T$ distribution of the dominant backgrounds for the (a) gluon-fusion and (b) $Z$-associated channels. Some example signals are also included (see Sec.~\ref{Sec:MassDetermination} for details).}
\label{fig:Backgrounds}
\end{center}
\end{figure}

A set of selection cuts are applied based on Ref.~\cite{CMS:2015ifd}. These are:
\begin{itemize}
  \item At least one photon with $p_T > 45~\text{GeV}$ and $|\eta| < 1.44$.
  \item $p_T^{\text{miss}} > 50$~GeV, where $p_T^{\text{miss}}$ is the norm of the missing transverse momentum.
  \item Less than two jets with $p_T > 30$~GeV and $|\eta| < 2.4$.
  \item No electrons with $p_T > 10$~GeV and $|\eta| < 1.44$ or $1.57 < |\eta| < 2.5$, where the omitted region corresponds to the barrel/endcap transition region.
  \item No muons with $p_T > 10$~GeV and $|\eta| < 2.1$, unless its angular distance $\Delta R = \sqrt{\Delta\eta^2 + \Delta\phi^2}$ from a valid jet is less than 0.3.
\end{itemize}
To validate our procedure, we reproduced the background estimates of Ref.~\cite{CMS:2015ifd} for their model-independent search at 8~TeV and using their exact cuts. We find good compatibility. The cuts at 14~TeV are adjusted to exploit the fact that the Higgs bosons produced in gluon-fusion have little transverse momentum. This leads to Jacobian edges in both the distributions of the $p_T$ of the photon and $p_T^{\text{miss}}$. Fig.~\ref{fig:BackGroundsGF} shows the distribution of the transverse mass $m_T$ for the different backgrounds at 14~TeV, where $m_T$ is defined as
\begin{equation}\label{eq:MTdef}
  m_T = \sqrt{2p_T^A p_T^{\text{miss}}(1 - \cos\Delta\phi(A, p_T^{\text{miss}}))},
\end{equation}
where $p_T^A$ is the transverse momentum of the leading photon and $\Delta\phi(A, p_T^{\text{miss}})$ is the difference between the azimuthal angles of the photon and the missing transverse momentum.\footnote{Do note that alternative definitions of the transverse mass exist \cite{Barr:2011xt}.} As can be seen, the dominant background around the mass of the Higgs boson comes from the mistagging of jets as photons. The peak in the background was at considerably lower $m_T$ for 8~TeV, but the increase in center-of-mass energy moved it to where the signal is expected for a massless dark photon.

\subsection{Backgrounds for $Z$-associated production}\label{sSec:ZassociatedBackgrounds}

The most important backgrounds for the $Z$-associated channel are presented in Table~\ref{table:Zassociated}, including the number of simulated events for each of them. 

A few points are worth mentioning. For the $WZ\;(e \to A)$ background, the $W$ is always decayed to an electron and a neutrino and the $Z$ is always decayed leptonically. This background only contains events in which an electron is mistagged as a photon. The reweighting is performed as in Sec.~\ref{sSec:GluonFusionBackgrounds}. Events that pass the cuts without mistagging are instead accounted in the $WZ$ background, which also includes the decay of the $W$ to a muon and a neutrino. For the $ZZ$ background, one of the $Z$'s is decayed to leptons and the other to neutrinos. For the $WW$ and $t\bar{t}$ backgrounds, the $W$'s are decayed leptonically to the same flavour. For the $ZA$ background, the $Z$ is decayed leptonically.

A set of selection cuts are applied based on Ref.~\cite{CMS:2019ajt}. They are:
\begin{itemize}
  \item Exactly two opposite-sign, same-flavour leptons, with $p_T > 25$ (20)~GeV for the leading (subleading) lepton and $|\eta| < 2.5$ (2.4) for electrons (muons).
  \item At least one photon with $p_T > 25$ GeV and $|\eta| < 2.5$.
  \item The invariant mass of the lepton pair $m_{\ell\ell}$ must satisfy $|m_{\ell\ell} - m_Z| < 15$ GeV, where $m_Z$ is the mass of the $Z$ boson.
  \item $p_T^{\text{miss}} > 110$~GeV.
  \item The norm of the vector sum of the transverse momenta of the leptons $p_T^{\ell\ell}$ must satisfy $p_T^{\ell\ell} > 60$ GeV.
  \item No jets tagged as originating from a bottom quark with $p_T > 20$~GeV and $|\eta| < 2.4$.
  \item Fewer than three jets with $p_T > 30$~GeV and $|\eta| < 4.7$.
  \item The difference in azimuthal angle between the lepton pair and the sum of $\vec{p}_T^{\text{miss}}$ and the momentum of the photon $\Delta\phi_{\vec{\ell\ell}, \vec{p}_T^{\text{miss}} + \vec{p}^A_T}$ must be more than 2.5~rad.
  \item $|p_T^{\vec{p}_T^{\text{miss}} + \vec{p}^A_T} - p_T^{\ell\ell}|/p_T^{\ell\ell} < 0.4$.
  \item The difference in azimuthal angle between the leading jet and $\vec{p}_T^{\text{miss}}$ must be larger than 0.5~rad.
  \item The invariant mass of the photon and the two leptons must be larger than 100~GeV.
  \item $m_T < 350$~GeV.
\end{itemize}
To validate our procedure, we reproduced the background estimates of Ref.~\cite{CMS:2019ajt} at 13~TeV. We find good compatibility with their results. The distribution of $m_T$ of the different backgrounds is shown in Fig.~\ref{fig:BackGroundsZH}.

\section{Mass determination}\label{Sec:MassDetermination}

We present in this section prospects for the upper limit on the mass of the dark photon in the massless case and uncertainties on the mass for the massive case. This is done for both gluon-fusion and $Z$-associated Higgs production. The dark photon is referred to as $A'$.\footnote{See Refs.~\cite{Gabrielli:2014oya, Biswas:2016jsh} for a discussion of the $h \to AA'$ signal at hadron colliders.}

\subsection{Upper limit on the mass of a massless dark photon}\label{sSec:MassLimitMassless}

\begin{figure}[t!]
\begin{center}
 \begin{subfigure}{0.49\textwidth}
    \centering
    \caption{Gluon-fusion}
    \includegraphics[width=1\textwidth]{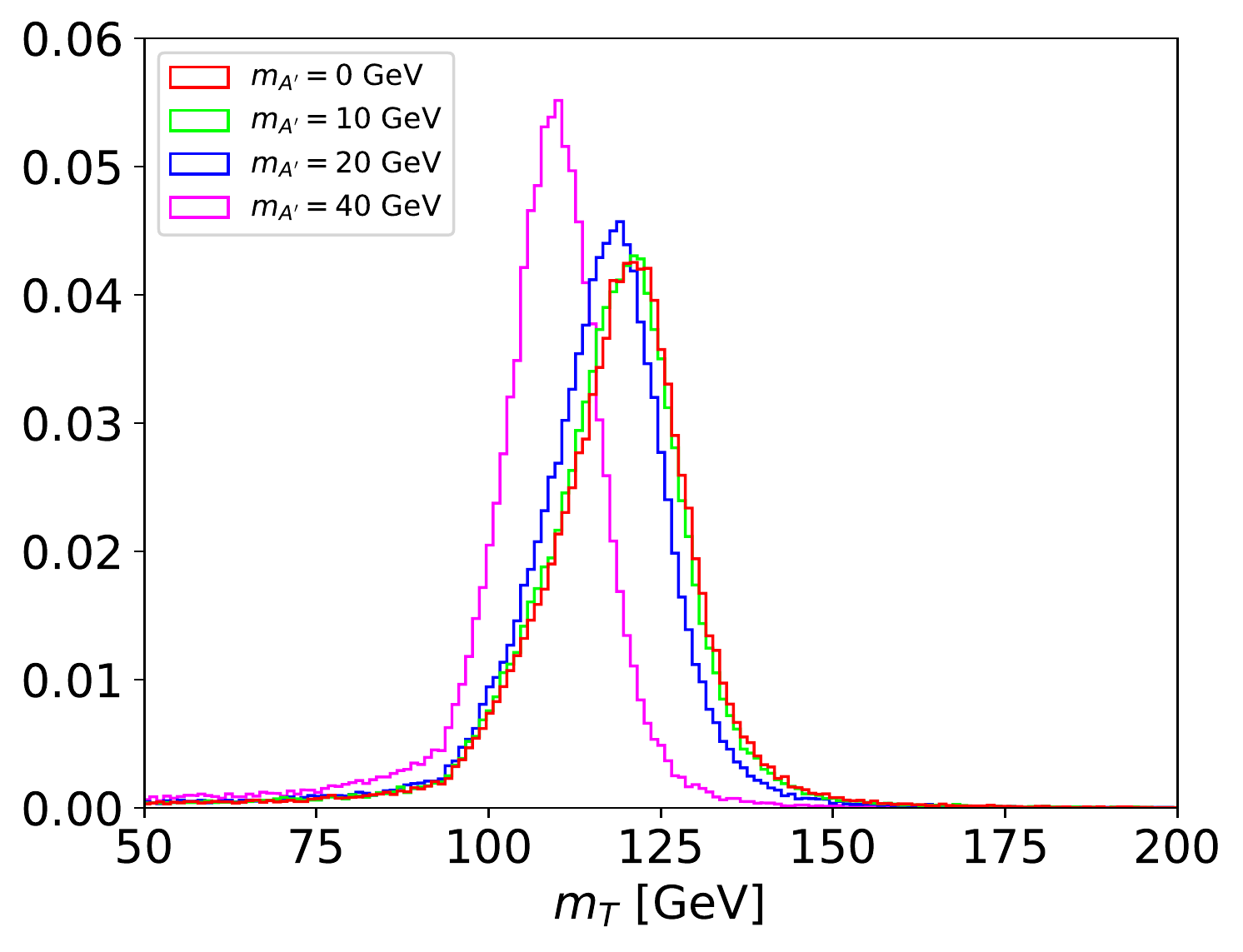}
    \label{fig:TemplatesGGF}
  \end{subfigure}
 \begin{subfigure}{0.49\textwidth}
    \centering
    \caption{$Z$-associated}
    \includegraphics[width=1\textwidth]{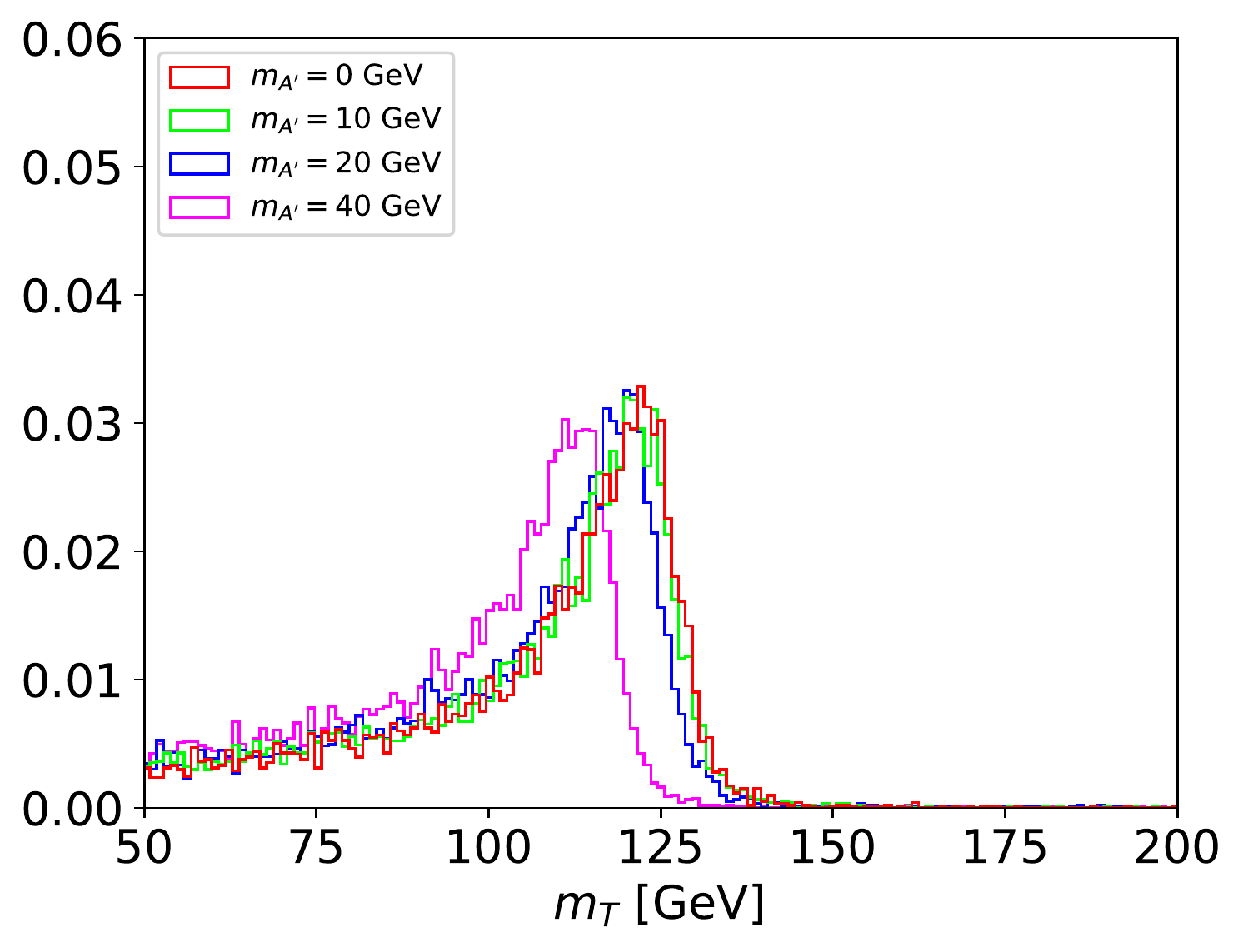}
    \label{fig:TemplatesZH}
  \end{subfigure}
\caption{Example templates of the $m_T$ distribution of the signal for (a) the gluon-fusion and (b) $Z$-associated channels. All histograms are normalized to 1.}
\label{fig:Templates}
\end{center}
\end{figure}

Upper limits on the mass of the dark photon for the massless case are obtained using likelihood methods. In more details, a series of templates of the transverse mass distribution with 1~GeV bin width are produced for different masses of the dark photon $m_{A'}$.\footnote{A dark photon heavier than a few GeV that is stable on collider scales might be difficult to justify. However, scenarios in which the dark photon decays almost exclusively to invisible stable particles are easy to conceive. Such scenarios would lead to identical kinematic distributions (up to very small width effects) and could not realistically be distinguished from a stable dark photon via this analysis alone.} For each template, $5\times 10^5$ signal events are generated and the cuts of Sec.~\ref{sSec:GluonFusionBackgrounds} or Sec.~\ref{sSec:ZassociatedBackgrounds} are applied. Some example templates are shown in Fig.~\ref{fig:Templates} for both Higgs production channels. The gluon-fusion and $Z$-associated cross sections are taken from Ref.~\cite{LHCHiggsCrossSectionWorkingGroup:2016ypw}. For a given integrated luminosity $\int L dt$ and branching ratio of the Higgs to a photon and a dark photon $\text{BR}(h\to AA')$, a series of toy experiments is performed using the template corresponding to the massless dark photon and fluctuating the number of events in each bin according to Poisson distributions. For each toy experiment, the likelihood of every mass template is computed using the bins from 80 to 140~GeV and assuming the same $\text{BR}(h\to AA')$. This produces the likelihood function $\mathcal{L}(m_{A'})$. The latter can then be used to obtain an upper limit on the mass of the dark photon. Considering that the likelihood function might not be well approximated by a normal distribution, we consider two statistical methods for this. First, a simple $\chi^2$ fit is performed by finding the mass with the highest likelihood, $m_{A'}^\text{max}$. The mass range allowed at 95\% CL is the one where $-2\ln\mathcal{L}(m_{A'}) + 2\ln\mathcal{L}(m_{A'}^\text{max}) < 3.84$ and the upper limit on the mass of the dark photon is the largest $m_{A'}$ in this range. Second, Bayesian statistics are used. Assuming a flat prior, the likelihood function corresponds to the posterior distribution up to a normalization constant. The range of non-excluded $m_{A'}$ at 95\% CL corresponds to the Highest Posterior Density (HPD) credible region with probability content 0.95, where the HPD credible region is the region with all density probabilities being higher than outside of it~\cite{Loredo:2001rx}. The upper limit on the mass of the dark photon is then the highest $m_{A'}$ in the HPD interval. By generating a sufficient number of toy experiments, a distribution of upper limits on $m_{A'}$ is obtained and its median value corresponds to the median expected upper limit on the mass of the dark photon.

To have an indication of where it would be possible to actually discover a dark photon, a signal region is defined by applying the cuts of Sec.~\ref{sSec:GluonFusionBackgrounds} or Sec.~\ref{sSec:ZassociatedBackgrounds} and also requiring $100\;\text{GeV} < m_T < 130\;\text{GeV}$. The region considered discoverable is where the significance $s/\sqrt{b}$ is larger than 5, where $s$ is the expected number of signals and $b$ the expected number of backgrounds. As an example, for an integrated luminosity of $139~\text{fb}^{-1}$ and a $\text{BR}(h\to AA')$ of $1\%$, the signal is about $1.20 \times 10^4$ (1.00) and the background $7.46 \times 10^6$ (0.94) for gluon-fusion ($Z$-associated). This gives a $s/\sqrt{b}$ of $\sim 4.39$ ($1.04$) for gluon-fusion ($Z$-associated), which can easily be rescaled to other $\text{BR}(h\to AA')$ and integrated luminosities. In addition, the current strongest constraints on $\text{BR}(h\to AA')$ is 1.8\% at 95\% CL, which comes from vector boson fusion with an integrated luminosity of $139~\text{fb}^{-1}$ and a center-of-mass energy of 13~TeV \cite{ATLAS:2021pdg}.

\begin{figure}[t!]
\begin{center}
 \begin{subfigure}{0.49\textwidth}
    \centering
    \caption{Gluon-fusion, $\chi^2$}
    \includegraphics[width=1\textwidth]{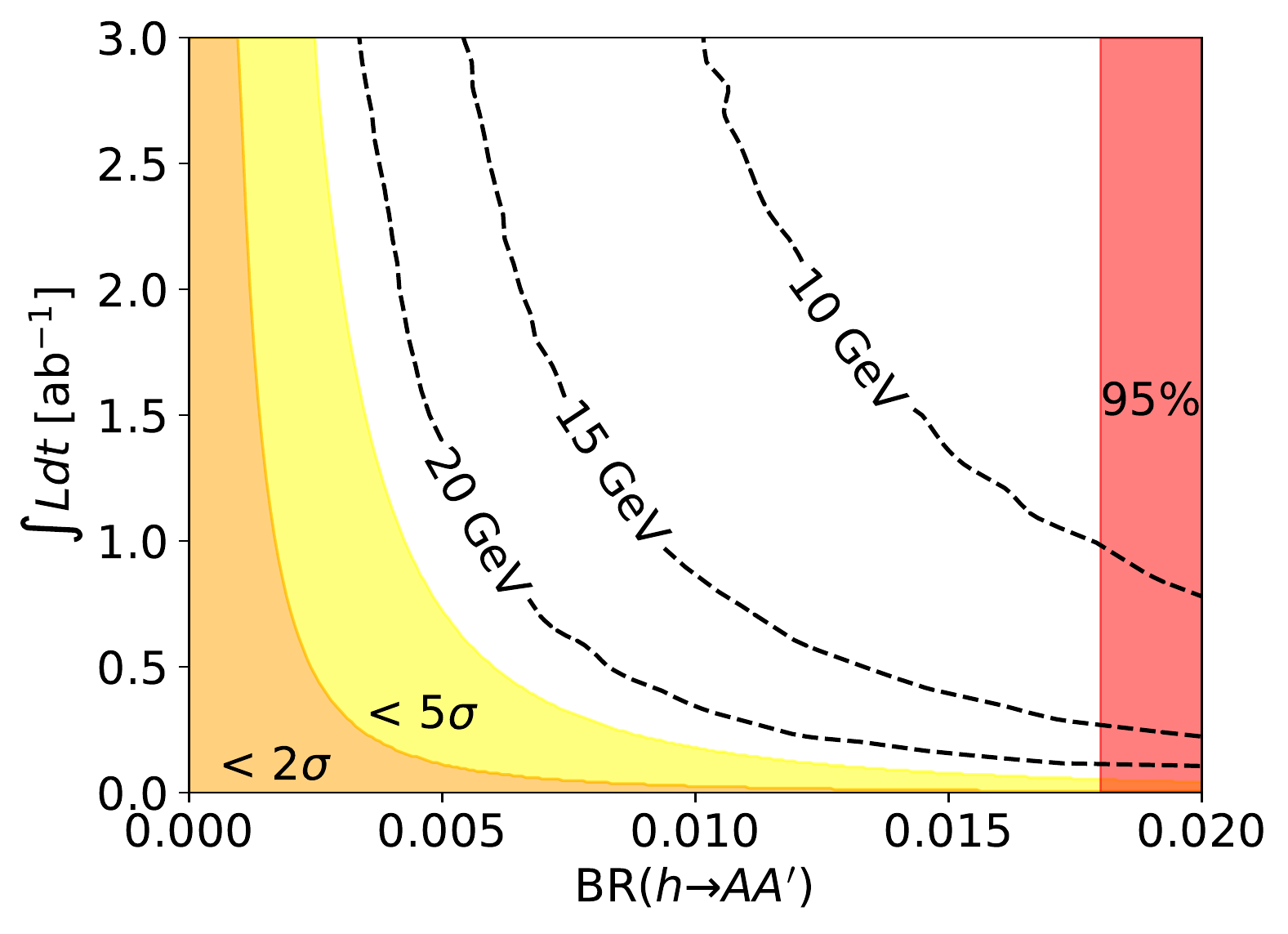}
    \label{fig:MaxMassChi2ggF}
  \end{subfigure}
 \begin{subfigure}{0.49\textwidth}
    \centering
    \caption{Gluon-fusion, Bayesian}
    \includegraphics[width=1\textwidth]{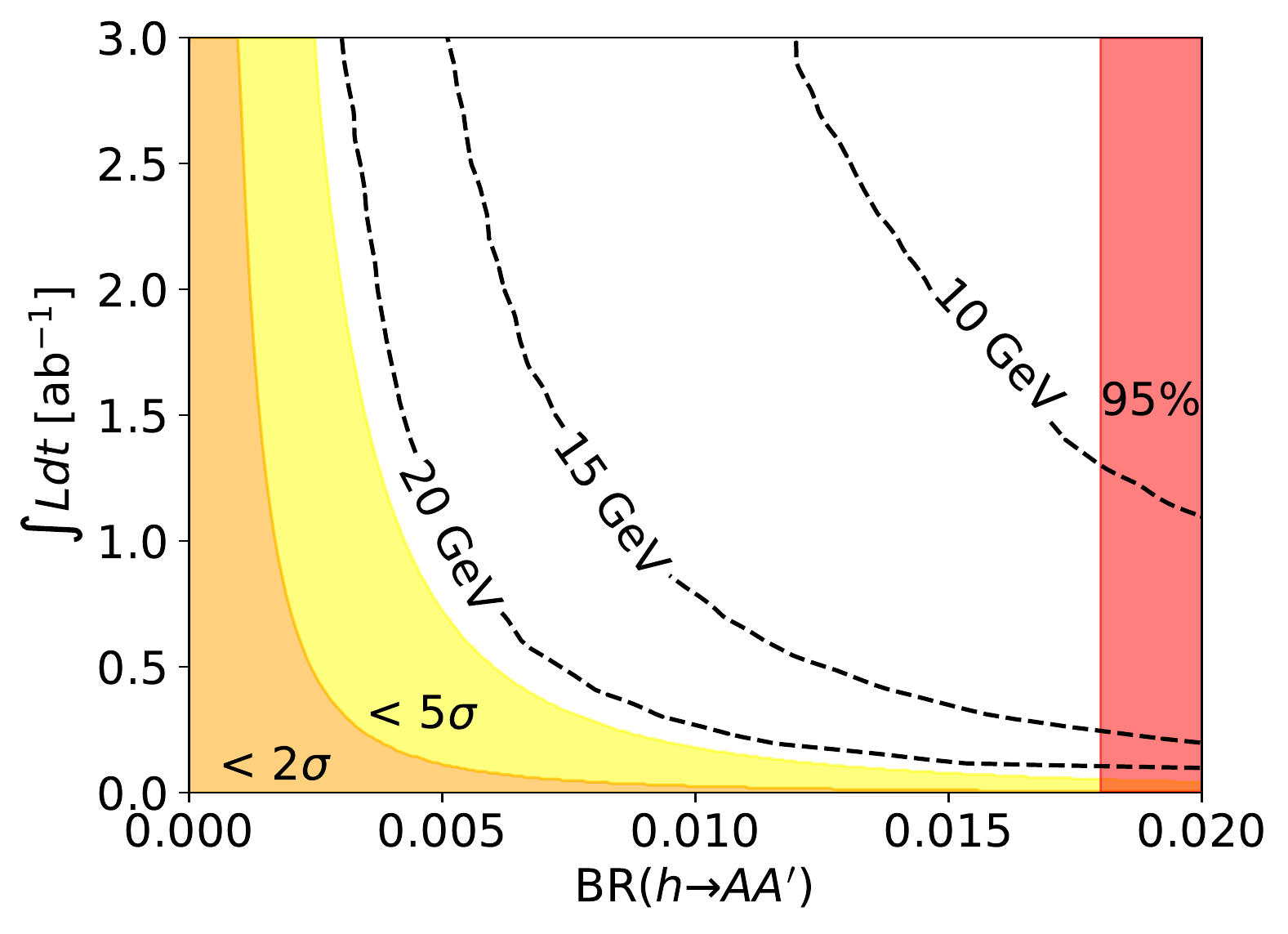}
    \label{fig:MaxMassBayeggF}
  \end{subfigure}
 \begin{subfigure}{0.49\textwidth}
    \centering
    \caption{$Z$-associated, $\chi^2$}
    \includegraphics[width=1\textwidth]{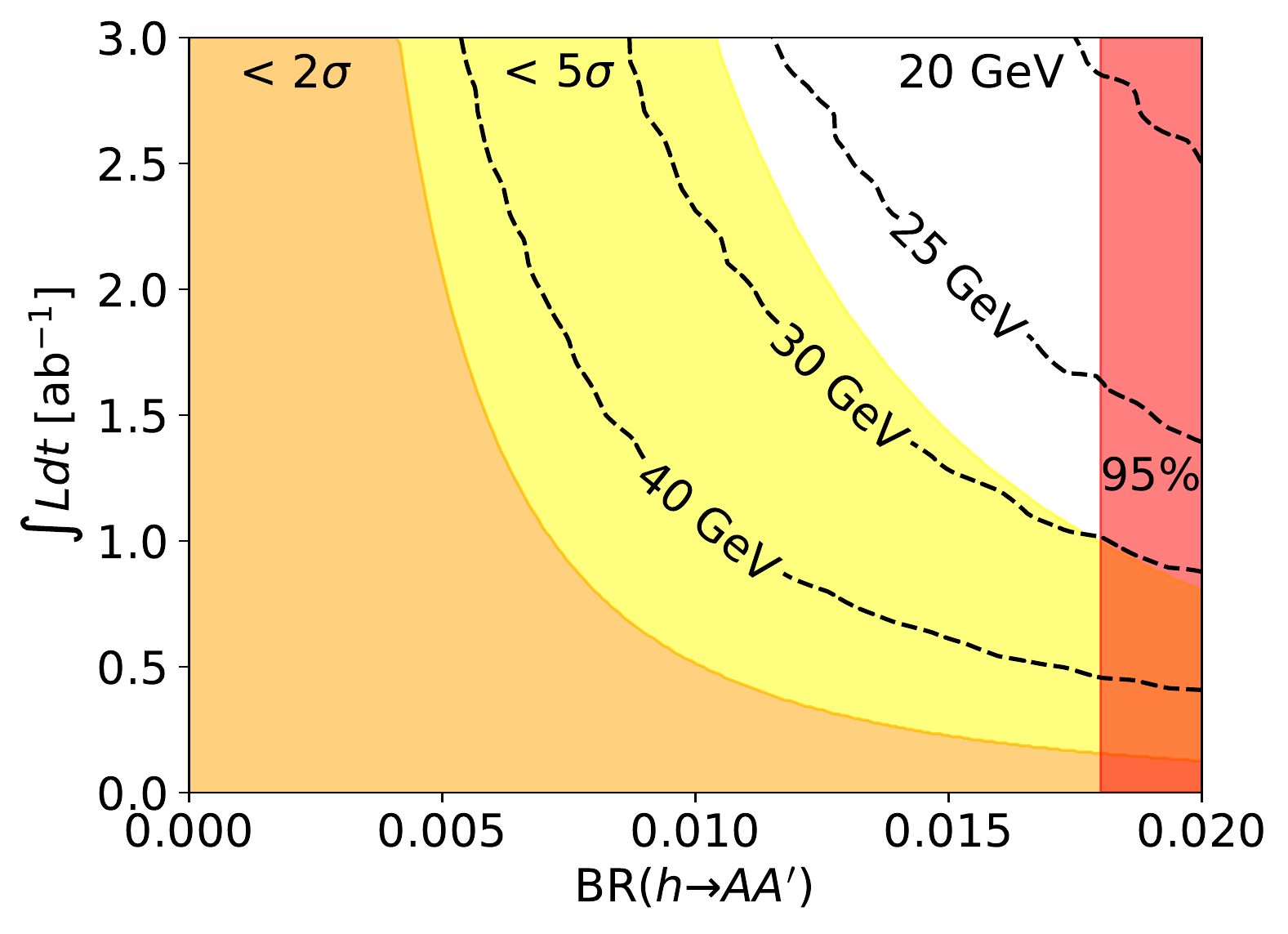}
    \label{fig:MaxMassChi2ZH}
  \end{subfigure}
 \begin{subfigure}{0.49\textwidth}
    \centering
    \caption{$Z$-associated, Bayesian}
    \includegraphics[width=1\textwidth]{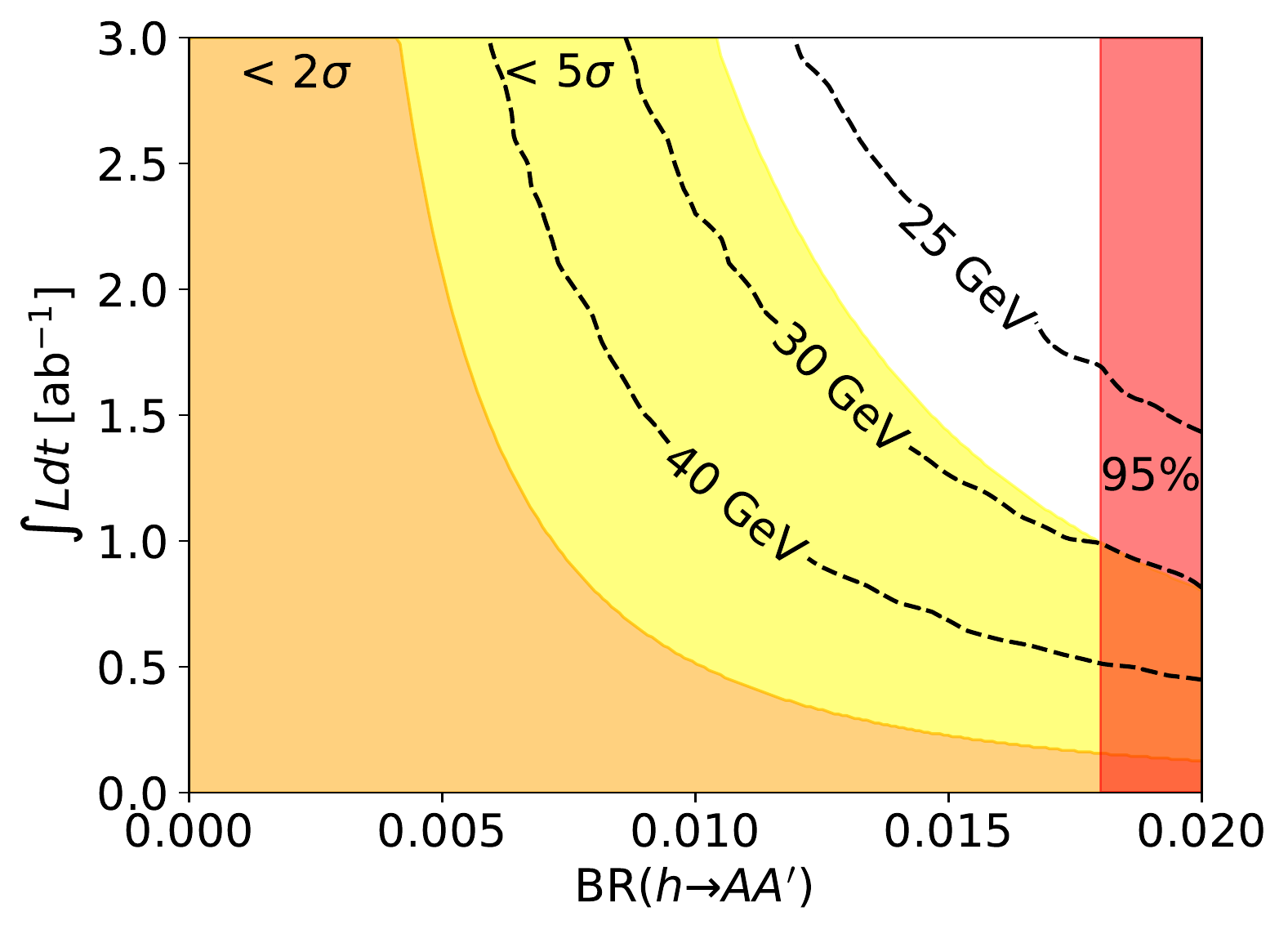}
    \label{fig:MaxMassBayeZH}
  \end{subfigure}
\caption{Upper limit at 95\% CL on the mass of a massless dark photon for different Higgs production channels and different statistical methods. The signal in the yellow (orange) region would have a significance of less than $5\sigma$ ($2\sigma$). The pink region is already excluded by current collider searches \cite{ATLAS:2021pdg}.}
\label{fig:MaxMass}
\end{center}
\end{figure}

The median expected upper limit on $m_{A'}$ is shown for gluon-fusion and $Z$-associated production in Fig.~\ref{fig:MaxMass} as a function of $\text{BR}(h\to AA')$ and the integrated luminosity for both the $\chi^2$ approach and the Bayesian approach.\footnote{The results are presented in terms of $\text{BR}(h\to AA')$ for the sake of model independence and can easily be applied to specific models. For kinetic mixing of the weak hypercharge with an Abelian gauge boson, $\text{BR}(h\to AA')$ would depend on the mixing coefficient.}
As can be seen, both statistical methods give similar results. The reason the $\chi^2$ limits are slightly stronger is the lower limit on the dark photon mass of 0, which makes the likelihood function sometimes differ considerably from a normal distribution. In addition, the limits on $m_{A'}$ could at most reach $\mathcal{O}(1)~\text{GeV}$. This is due to the fact that the Jacobian edge in the $m_T$ distribution is around
\begin{equation}\label{eq:mTJacobianEdge}
  m_T^\text{max} = m_h\left(1 - \frac{m_{A'}^2}{m_h^2}\right) = m_h - \frac{m_{A'}^2}{m_h},
\end{equation}
where $m_h$ is the mass of the Higgs boson. This means that the Jacobian edge is only displaced from $m_h$ by a term quadratic in $m_{A'}$ and hence why small masses of the dark photon are difficult to constrain. Finally, the gluon-fusion channel gives considerably stronger limits than the $Z$-associated channel. The main reason is that the signal would be easier to distinguish from the background for the gluon-fusion channel. This is compounded by the fact that the background for the $Z$-associated channel is very small. Hence, a discovery could be made with very few events. Having few events however makes it difficult to tell apart different hypotheses.

\subsection{Uncertainty on the mass of a massive dark photon}\label{sSec:MassUncertaintyMassive}

In the case of a massive dark photon, an uncertainty can be set on its mass. This can be done with minimal modifications of either the $\chi^2$ or Bayesian approaches. In both cases, pseudo-experiments are generated as before, but now using a template for a dark photon with a given non-zero mass. For each pseudo-experiment, the unexcluded region at $1\sigma$ is found. This defines both an upper limit and a lower limit on the mass of the dark photon. We take the uncertainty to be the difference between these two numbers divided by 2. Generating many pseudo-experiments defines a distribution of uncertainties and its median defines the median expected uncertainty on the mass of the dark photon.

Similar to Sec.~\ref{sSec:MassLimitMassless}, a signal region can be defined to determine the region where the signal is discoverable. The only difference is that the $m_T$ requirement is generalized to $m_T \in \left[m_T^\text{max} - 25~\text{GeV},  m_T^\text{max} + 5~\text{GeV}\right]$. For a dark photon considerably heavier than what we consider, it would be necessary to reoptimize the selection cuts of Sec.~\ref{Sec:BackgroundSimulation}, but this is beyond the scope of this paper.

\begin{figure}[t!]
\begin{center}
 \begin{subfigure}{0.49\textwidth}
    \centering
    \caption{Gluon-fusion, $m_{A'}=10~\text{GeV}$}
    \includegraphics[width=1\textwidth]{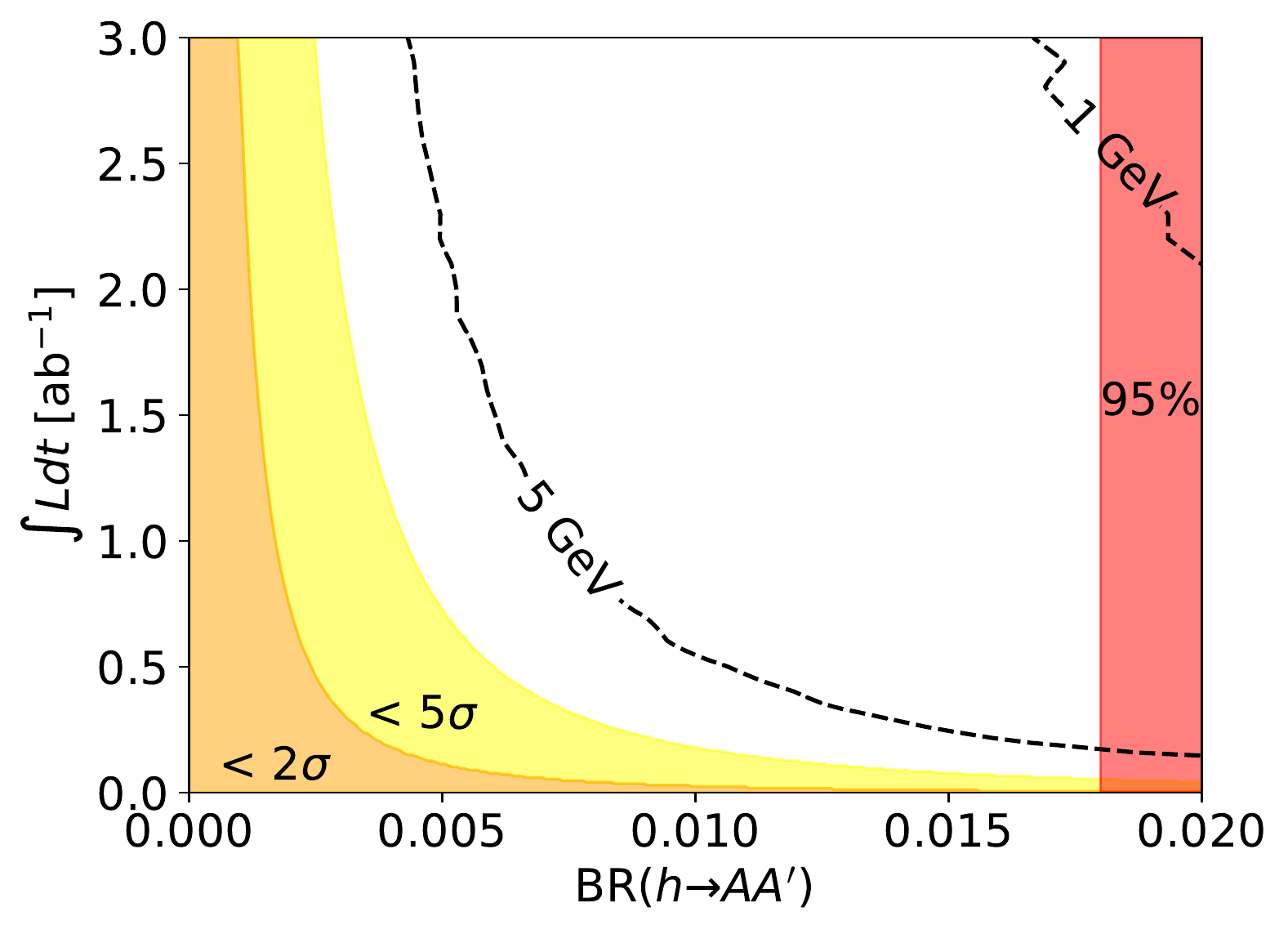}
    \label{fig:UncChi2ggF_10}
  \end{subfigure}
 \begin{subfigure}{0.49\textwidth}
    \centering
    \caption{Gluon-fusion, $m_{A'}=20~\text{GeV}$}
    \includegraphics[width=1\textwidth]{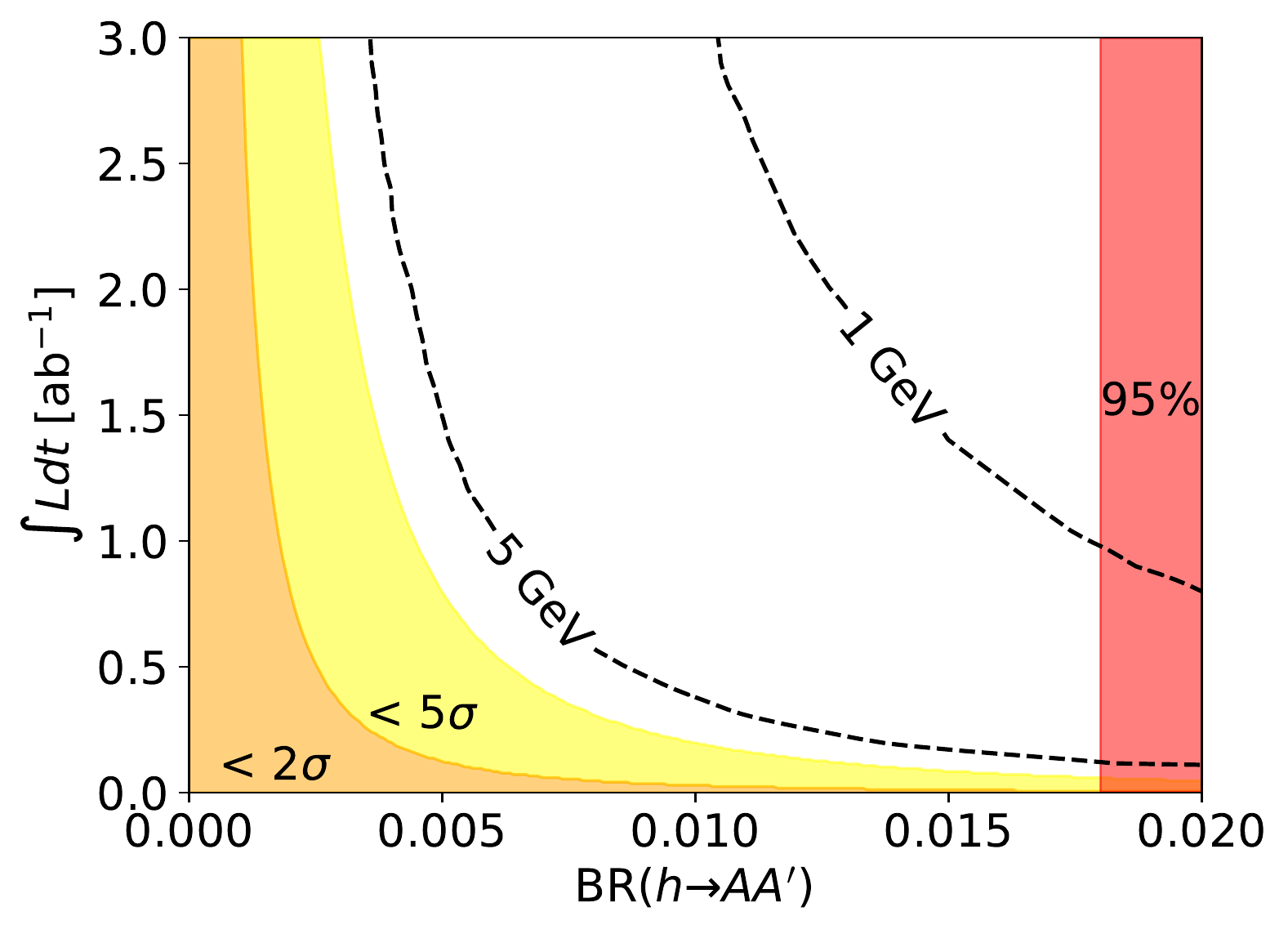}
    \label{fig:UncChi2ggF_20}
  \end{subfigure}
   \begin{subfigure}{0.49\textwidth}
    \centering
    \caption{$Z$-associated, $m_{A'}=10~\text{GeV}$}
    \includegraphics[width=1\textwidth]{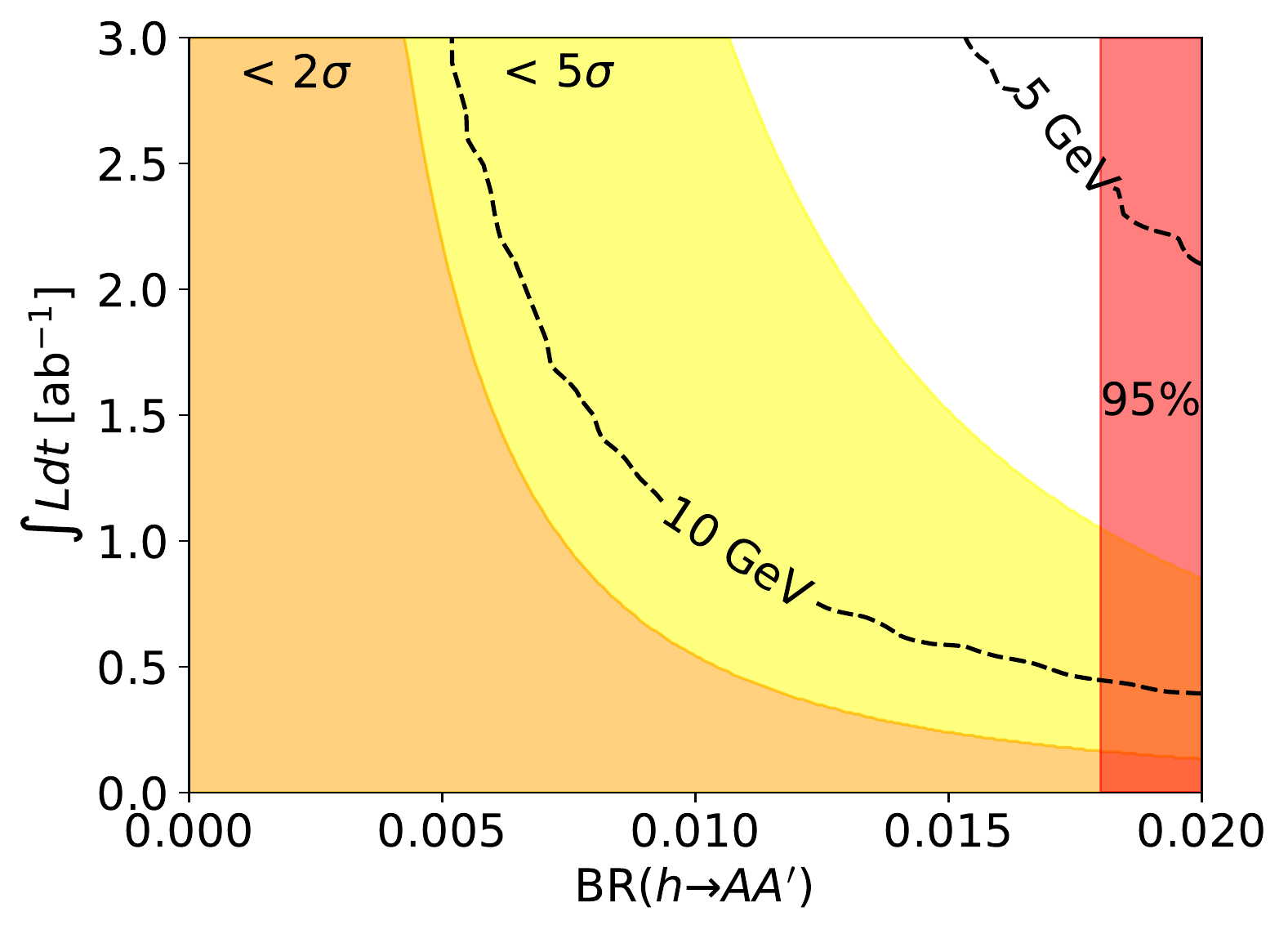}
    \label{fig:UncChi2ZH_10}
  \end{subfigure}
 \begin{subfigure}{0.49\textwidth}
    \centering
    \caption{$Z$-associated, $m_{A'}=20~\text{GeV}$}
    \includegraphics[width=1\textwidth]{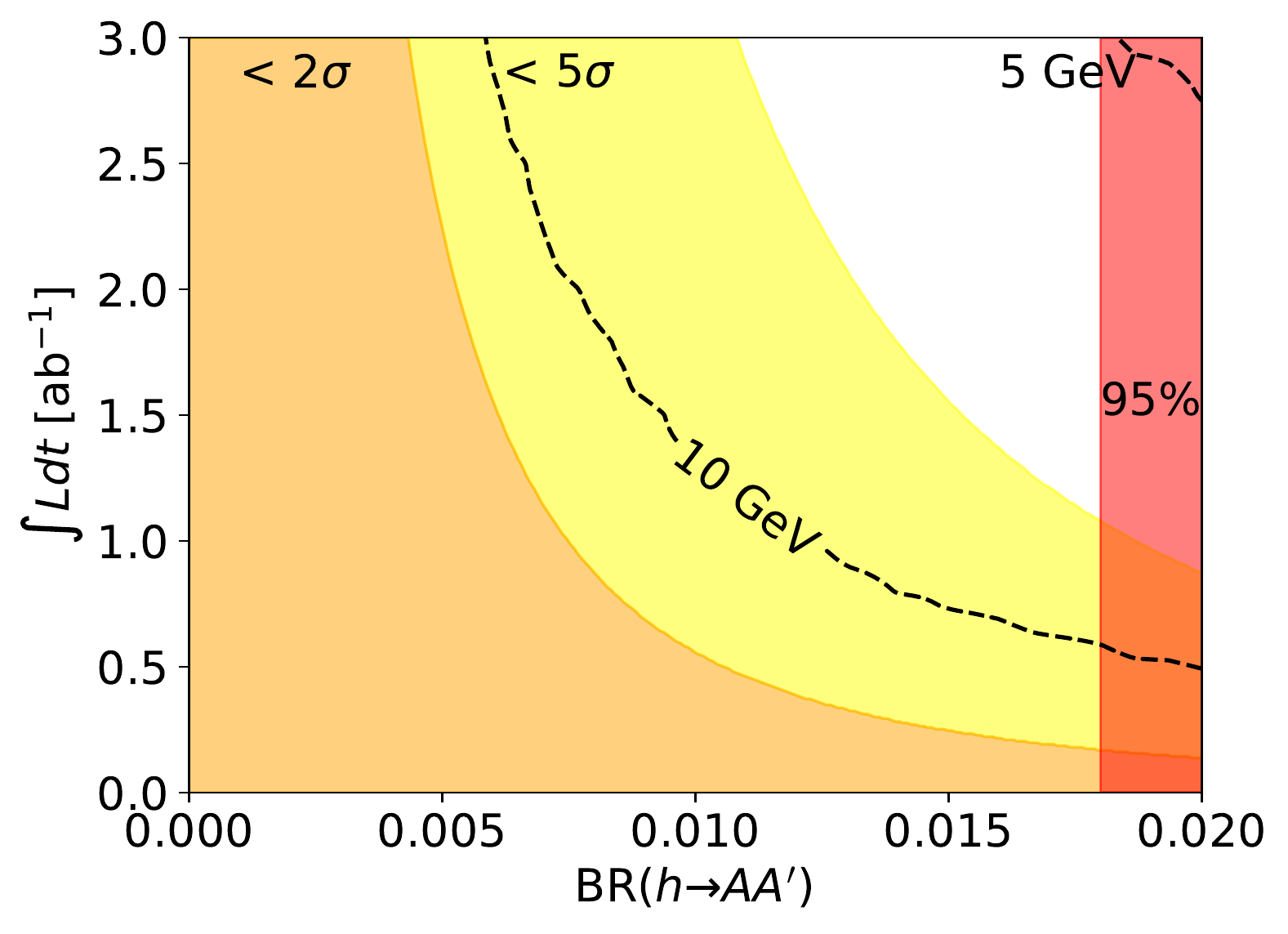}
    \label{fig:UncChi2ZH_20}
  \end{subfigure}
\caption{Uncertainty on the mass of a massive dark photon for the different Higgs production channels. The signal in the yellow (orange) region would have a significance of less than $5\sigma$ ($2\sigma$).  The pink region is already excluded by current collider searches \cite{ATLAS:2021pdg}. The limits are obtained using the $\chi^2$ method.}
\label{fig:UncMass}
\end{center}
\end{figure}

The median expected uncertainty on $m_{A'}$ is shown in Fig.~\ref{fig:UncMass} for the $\chi^2$ approach, different production channels and different $m_{A'}$. The results for the Bayesian approach are qualitatively similar. The gluon-fusion channel once again proves to give the more precise results because of the higher number of events. The precision with which the mass of the dark photon can be measured is controlled by different competing effects. On one hand, Eq.~\eqref{eq:mTJacobianEdge} means that the position of the edge is very insensitive to the value of $m_{A'}$ when it is small. On the other hand, a dark photon too close in mass to the Higgs boson is either less likely to pass the selection cuts (gluon-fusion) or will have its edge in a region with larger backgrounds ($Z$-associated), both rendering the mass measurement more difficult.

\section{Distinguishing a dark photon from the neutralino/gravitino hypothesis}\label{Sec:Sec:N1N2}

Finding an excess in a search for the Higgs boson decaying to a dark photon and a photon does not necessarily mean that a dark photon has been discovered. In this section, we present how the decay of the Higgs boson to a photon and a dark photon could be distinguished from another well-motivated hypothesis.

Supersymmetry (SUSY) can potentially solve the hierarchy problem, provide a viable dark matter candidate and lead to gauge coupling unification, among others. One of its most sought-after signature is the decay of the Higgs boson to a light gravitino $G$ and a neutralino $N$, with the neutralino decaying to a photon and another gravitino. This scenario has been the subject of both theoretical (e.g., Refs~\cite{Djouadi:1997gw, Petersson:2012dp}) and experimental studies (e.g., Ref.~\cite{CMS:2015ifd}).\footnote{A light gravitino is common in models of gauge mediation (see Ref.~\cite{Giudice:1998bp} for a review). However, the absence of tachyonic states would require a SUSY breaking scale that makes it difficult to obtain a sizable $\text{BR}(h\to AA')$ \cite{Petersson:2012dp}. Nonetheless, there exists many SUSY mediation mechanism where this branching ratio can be sizable, as shown in Ref.~\cite{Petersson:2012dp}, and exploring every one of them is beyond the scope of this paper.} Most importantly, this SUSY signature is very similar to the Higgs boson decaying to a photon and a dark photon. This is especially true if the neutralino is not much lighter than the Higgs boson. In this case, the neutralino is produced with momentum nearly identical to that of the Higgs boson and the first gravitino is very soft. The neutralino then decays and transmits about half of its center-of-mass energy to the photon and the other half to the second gravitino. The final state then contains both a photon and an invisible particle with an energy in the Higgs center-of-mass frame of nearly half the mass of the Higgs boson, which closely mimics the dark photon scenario

The relevant interactions for this SUSY scenario can be encoded in the Lagrangian~\cite{Petersson:2012dp}
\begin{equation}\label{eq:LagrangianN1N2}
  \mathcal{L} = \hat{g} h \bar{G} N + \frac{1}{\Lambda}A_{\mu\nu}\bar{G}\sigma^{\mu\nu}N,
\end{equation}
where $\hat{g}$ is some coupling constant, $\Lambda$ is some scale, $A_{\mu\nu}$ is the field strength of the photon and $\sigma^{\mu\nu}=\frac{i}{2}[\gamma^\mu,\gamma^\nu]$. We have assumed $G$ and $N$ to be Majorana fermions.

To determine how well the dark photon hypothesis could be distinguished from the decay of a Higgs to a neutralino and a gravitino, we proceed as follows. A series of templates of the $m_T$ distribution for the SUSY hypothesis are generated for different masses of $N$ up to the mass of the Higgs boson. Consider one such template. For given integrated luminosity and $\text{BR}(h\to AA')$, pseudo-experiments are generated using the template for a massless dark photon. The ratio of the likelihood of the dark photon and SUSY hypotheses is then computed, assuming the worst case scenario that the average number of events that pass selection cuts are the same and using the bins from 80 to 140~GeV. By generating enough pseudo-experiments, we obtain a distribution of the likelihood ratio for the dark photon hypothesis. The procedure is then repeated for the SUSY hypothesis, obtaining a distribution of the likelihood ratio assuming the SUSY hypothesis. One can then compute the p-value of the median of the dark photon distribution for the SUSY hypothesis. This gives the median expected p-value. The procedure is then repeated for other masses of $N$.

\begin{figure}[t!]
\begin{center}
 \begin{subfigure}{0.49\textwidth}
    \centering
    \caption{Gluon-fusion}
    \includegraphics[width=1\textwidth]{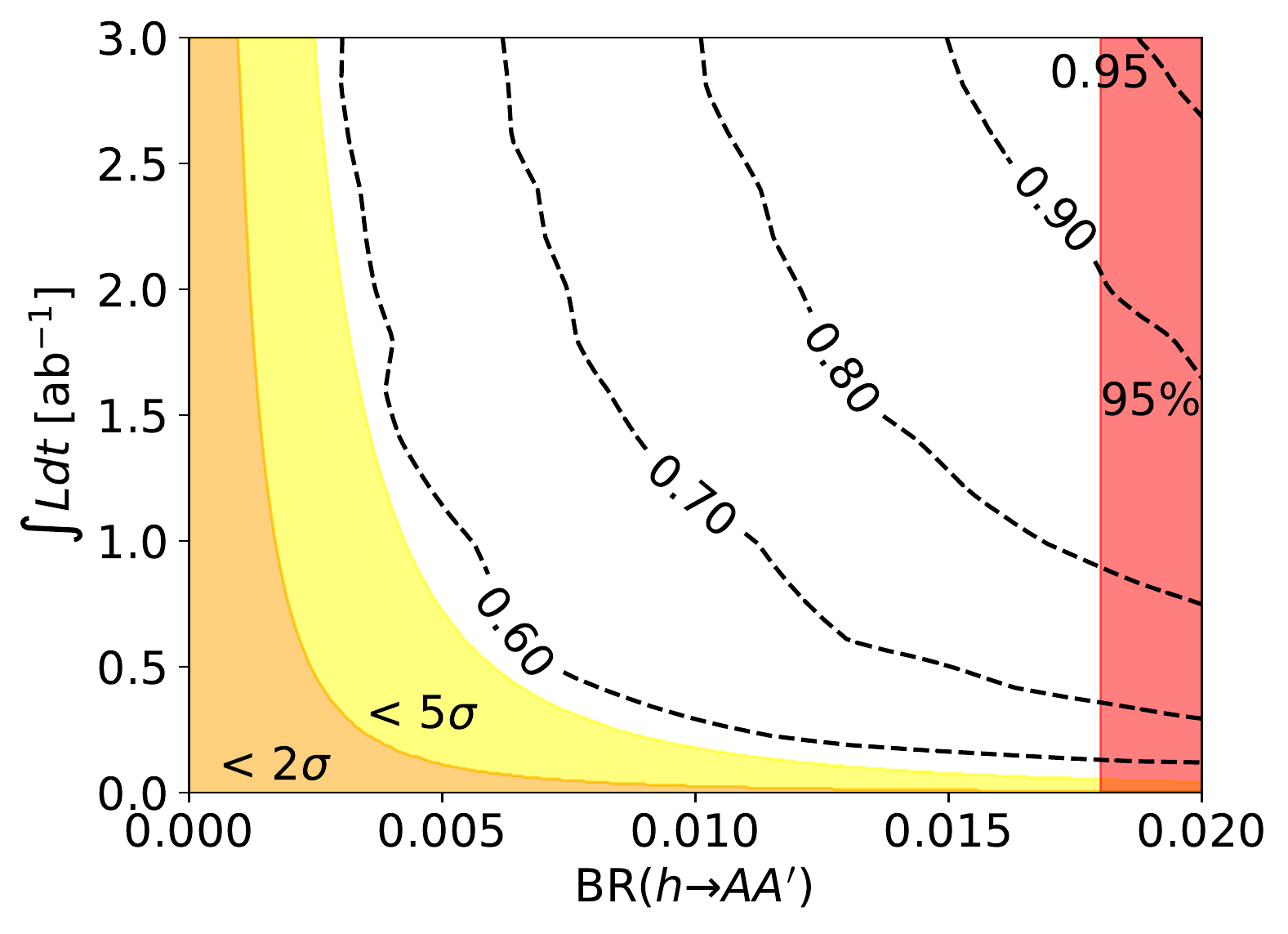}
    \label{fig:PvalueN1N2ggF}
  \end{subfigure}
 \begin{subfigure}{0.49\textwidth}
    \centering
    \caption{$Z$-associated}
    \includegraphics[width=1\textwidth]{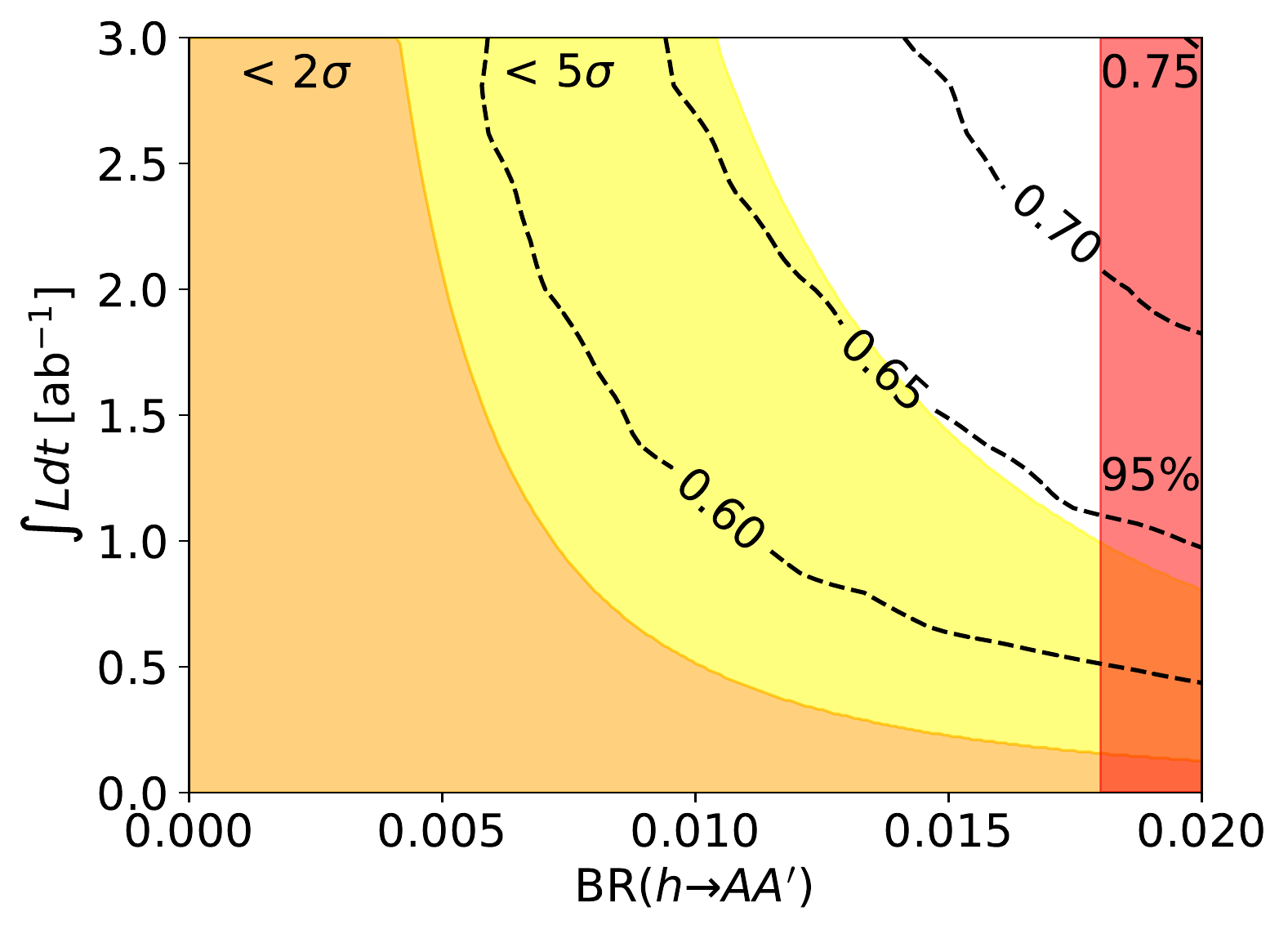}
    \label{fig:PvalueN1N2ZH}
  \end{subfigure}
\caption{Lowest exclusion confidence level of the SUSY hypothesis for (a) the gluon-fusion and (b) $Z$-associated channels. The signal in the yellow (orange) region would have a significance of less than $5\sigma$ ($2\sigma$).  The pink region is already excluded by current collider searches \cite{ATLAS:2021pdg}.}
\label{fig:N1N2pvalue}
\end{center}
\end{figure}

We show in Fig.~\ref{fig:N1N2pvalue} the lowest exclusion confidence level amongst all SUSY templates as a function of $\text{BR}(h\to AA')$ and the integrated luminosity for both the gluon-fusion and the $Z$-associated channels. In other words, a contour of $X\%$ means that the SUSY hypothesis is excluded at $X\%$ CL for the mass of the neutralino that mimics most closely the dark photon signal and is excluded with a higher confidence for all other masses. The width of $N$ is assumed small. As can be seen, the LHC could come close to ruling out the SUSY hypothesis at 95\% CL depending on the branching ratio and integrated luminosity. In addition, the gluon-fusion channel once again outperforms the $Z$-associated channel.

The fact that the SUSY hypothesis could be almost excluded might seem counterintuitive at first. Indeed, it might seem that, in the limit that the mass of $N$ approaches that of the Higgs boson, the SUSY and dark photon signals should become identical. One would naively expect this to be true until the mass splitting becomes comparable to the width of the Higgs, where the narrow width approximation breaks down and the Higgs boson is considerably off-shell in many events. This naive intuition proves to be not quite accurate. Consider for example gluon-fusion. The cross section for $g g \to N G$ is 
\begin{equation}\label{eq:ggtoNGCS}
  \sigma(\hat{s}) = \frac{\hat{g}^2}{16\pi\tilde{\Lambda}^2}\frac{(\hat{s} - m_N^2)^2}{(\hat{s} - m_h^2)^2 + m_h^2 \Gamma_h^2},
\end{equation}
where $\sqrt{\hat{s}}$ is the partonic center-of-mass energy, $\tilde{\Lambda}$ some effective scale, $m_N$ the mass of $N$ and $\Gamma_h$ the width of the Higgs boson. This can be rewritten in the more revealing form
\begin{equation}\label{eq:ggtoNGCS2}
  \sigma(\hat{s}) = \frac{\hat{g}^2}{16\pi\tilde{\Lambda}^2}\left[\frac{(\hat{s} - m_h^2)^2}{(\hat{s} - m_h^2)^2 + m_h^2 \Gamma_h^2} + \frac{2(m_h^2 - m_N^2)(\hat{s} - m_h^2)}{(\hat{s} - m_h^2)^2 + m_h^2 \Gamma_h^2} + \frac{(m_h^2 - m_N^2)^2}{(\hat{s} - m_h^2)^2 + m_h^2 \Gamma_h^2} \right].
\end{equation}
The first two terms are zero when the Higgs is on-shell. The last term represents the peak, though it is strongly suppressed because of the small mass splitting, and is small when the Higgs is off-shell. Call $f(\hat{s})d\hat{s}$ the probability for a collision to take place between $\hat{s}$ and $\hat{s} + d\hat{s}$. The Higgs will be mostly on-shell if
\begin{equation}\label{eq:OnOffshell1}
  \int_{(m_h - n\Gamma_h)^2}^{(m_h + n\Gamma_h)^2}d\hat{s} f(\hat{s})\sigma(\hat{s})\sim \frac{\hat{g}^2(m_h^2 - m_N^2)^2}{16\pi\tilde{\Lambda}^2m_h^3\Gamma_h},
\end{equation}
where $n$ is an integer of $\mathcal{O}(1)$, is much larger than
\begin{equation}\label{eq:OnOffshell2}
  \int_{(m_h + n\Gamma_h)^2}^{s}d\hat{s} f(\hat{s})\sigma(\hat{s})\sim \frac{\hat{g}^2}{16\pi\tilde{\Lambda}^2},
\end{equation}
where $\sqrt{s}$ is the center-of-mass energy of the hadrons and we have performed an order of magnitude estimate. This means that most events will take place when the Higgs is on-shell only when $m_h - m_N \gtrsim \sqrt{m_h \Gamma_h}$, which is much larger than the naive result. The validity of this relation was verified by modifying $\Gamma_h$ in \texttt{MadGraph} and our own computation of the cross section. This phenomenon is why there is an important limit to how much the SUSY signal can mimic the dark photon signal and why it would be possible to eventually exclude the SUSY hypothesis. It is also worth noting that the SUSY hypothesis would be even easier to distinguish from a dark photon if the neutralino had a sizable width, as it would often become off-shell which would further modify the $m_T$ distribution.

\section{Observability of the Higgs decay to a $Z$ and a dark photon}\label{Sec:HiggstoZgammaD}

If the Higgs boson can decay to a photon and a dark photon, it is very likely that it can also decay to a dark photon and a $Z$ boson. In the presence of an excess in the Higgs decay to a photon and an invisible particle, the absence of an excess in the decay to a $Z$ boson and an invisible particle could potentially put tension on the dark photon hypothesis. In this section, we discuss the relative size of the branching ratios of these two channels. We consider two possibilities for how the dark photon communicates with the Standard Model.

\begin{figure}[t!]
\begin{center}
 \begin{subfigure}{0.49\textwidth}
    \centering
    \caption{Kinetic mixing only}
    \includegraphics[width=1\textwidth]{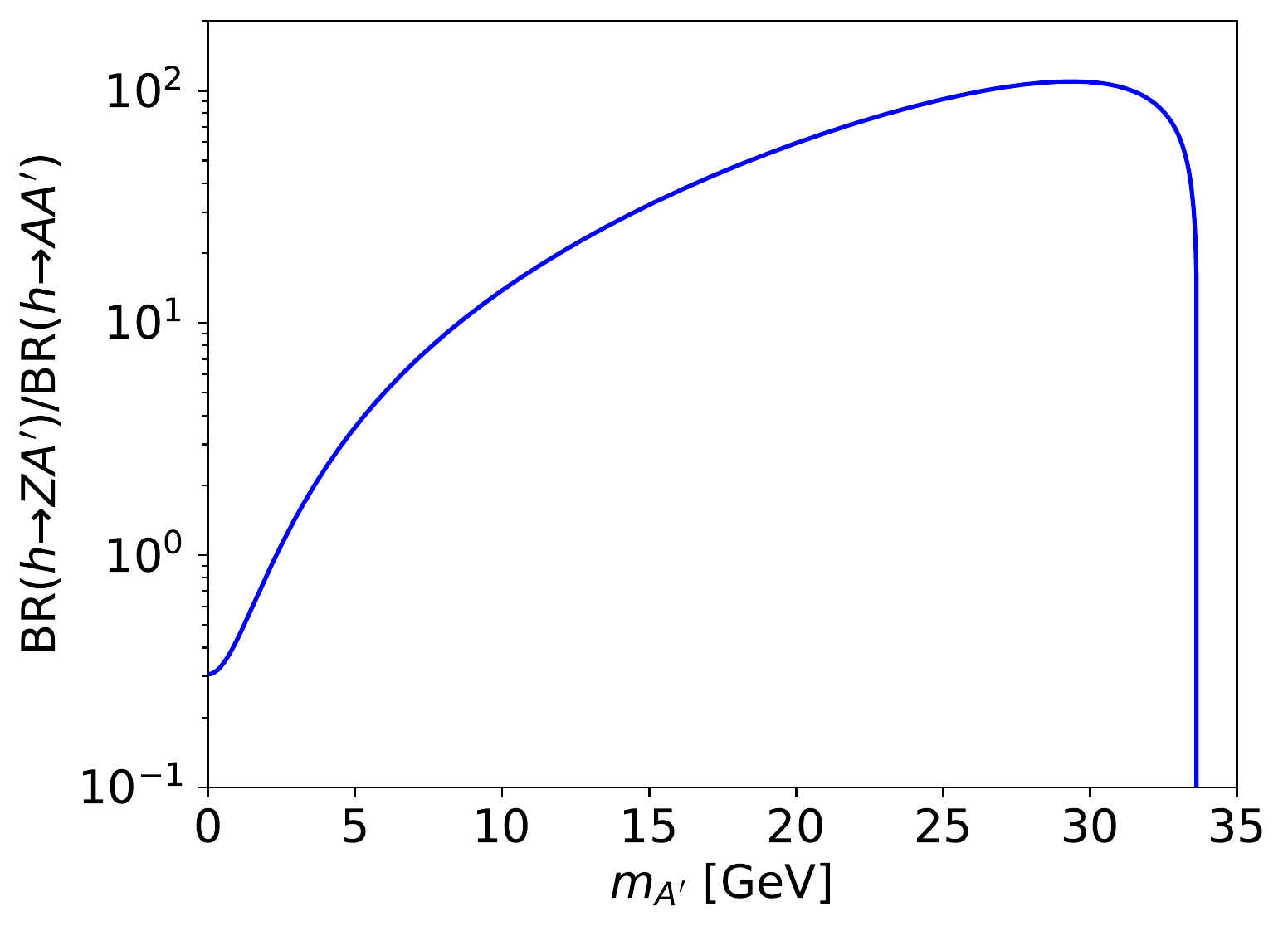}
    \label{fig:BRratio1}
  \end{subfigure}
 \begin{subfigure}{0.49\textwidth}
    \centering
    \caption{Effective operators}
    \includegraphics[width=1\textwidth]{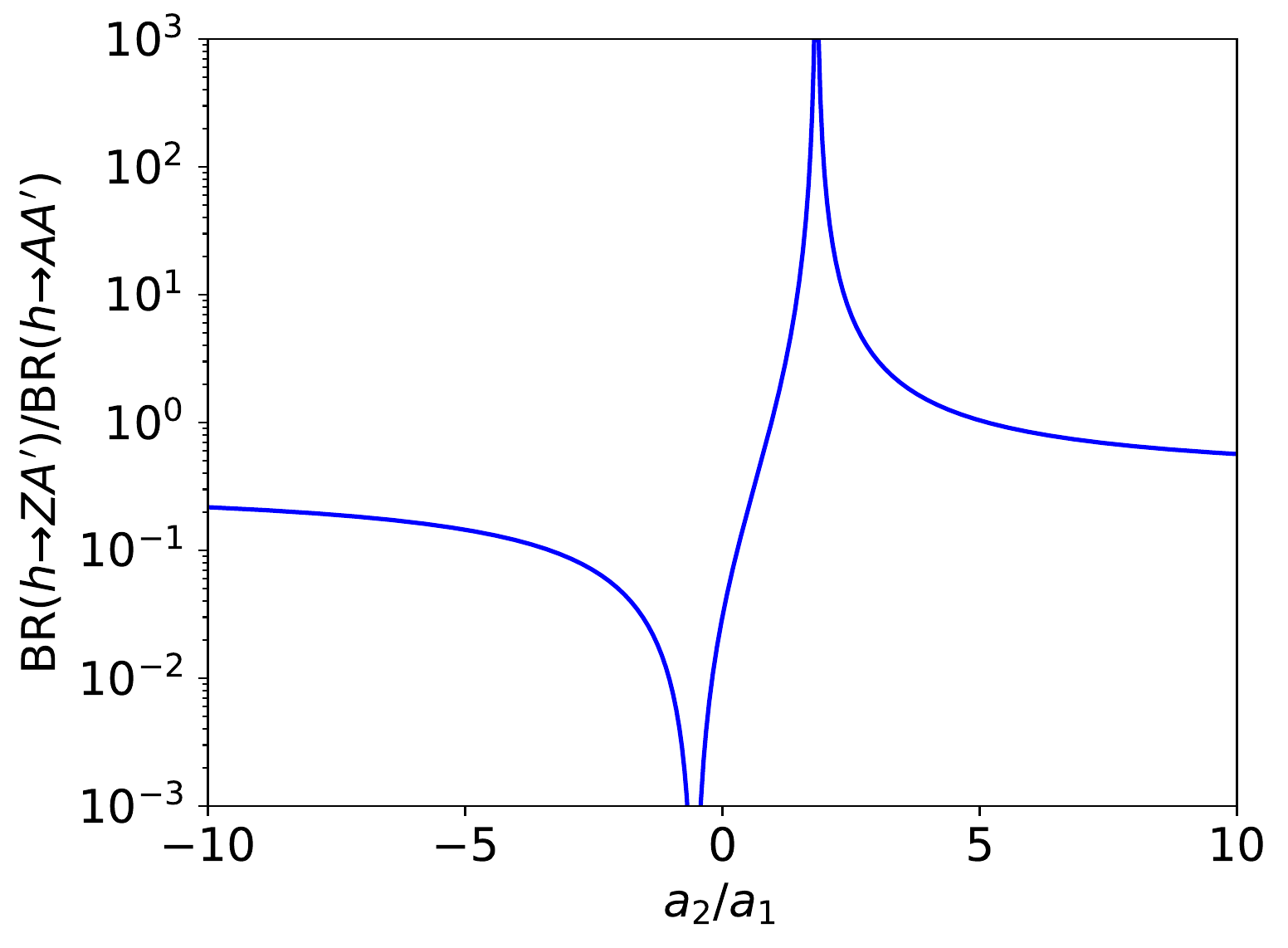}
    \label{fig:BRratio2}
  \end{subfigure}
\caption{Ratio of the branching ratios $\text{BR}(h \to Z A')/\text{BR}(h \to AA')$ for (a) kinetic mixing for a sufficiently small value of the mixing parameter and (b) the effective operators of Eq.~\eqref{eq:OperatorsDefinition} with a very light dark photon.}
\label{fig:BR}
\end{center}
\end{figure}

First, a new Abelian gauge boson $\hat{Z}_D$ could mix with the hypercharge via the kinetic mixing term
\begin{equation}\label{eq:Mixing}
  \frac{1}{2} \frac{\epsilon}{c_W}\hat{Z}_{D\mu\nu}\hat{B}^{\mu\nu},
\end{equation}
where $c_W$ ($s_W$) is the cosine (sine) of the Weinberg angle. The technical details are presented in appendix~\ref{Sec:DecayWidths}. In the limit of small $\epsilon$, the ratio of the branching ratios to $ZA'$ and $AA'$ becomes independent of $\epsilon$. It is shown in Fig.~\ref{fig:BRratio1} for a very small value of the mixing parameter. As can be seen, the branching ratio to a $Z$ can be by far dominant for a sufficiently heavy dark photon. This is because the decay can take place at tree-level when the dark photon is massive. Even if the dark photon is very light, it is still $\sim 0.31$ of the value of the branching ratio to a photon and a dark photon. As such, kinetic mixing only means that the decay of the Higgs boson to a $Z$ boson and a dark photon should eventually be observable if the decay is kinematically allowed. Do note however that this signal would be more challenging to discover due to the smaller amount of missing transverse momentum.

Second, we consider the possibility of new particles leading to Higgs decay via loop corrections. For illustration purposes, we consider the following effective operators
\begin{equation}\label{eq:OperatorsDefinition}
  \frac{a_1}{\Lambda^2}H^\dagger H B_{\mu\nu}{A'}^{\mu\nu}, \qquad \frac{a_2}{\Lambda^2}H^\dagger \sigma^a H W^a_{\mu\nu}{A'}^{\mu\nu},
\end{equation}
where $A'_{\mu\nu}$ is the field strength of the dark photon, $\Lambda$ some scale and $a_1$ and $a_2$ some constants. Computing the branching ratios at leading order in the mass of the dark photon and the inverse of $\Lambda_1$ and $\Lambda_2$, one obtains
\begin{equation}\label{eq:BRratio2}
  \frac{\text{BR}(h\to Z A')}{\text{BR}(h\to AA')} = \left(\frac{s_W + c_W \frac{a_2}{a_1}}{c_W - s_W \frac{a_2}{a_1}}\right)^2
  \left(1 - \frac{m_Z^2}{m_h^2}\right)^3.
\end{equation}
This ratio is shown in Fig.~\ref{fig:BRratio2}. As can be seen, the ratio can easily be much smaller or much larger than one. The inclusion of subleading terms does not change the qualitative picture. As such, the absence of observation of the Higgs decay to a photon and a $Z$ boson would be perfectly compatible with new particles mediating the decay of the Higgs boson to a dark photon and a photon.

\section{Conclusion}\label{Sec:Conclusion}

The goal of this paper was to determine which properties of the dark photon could be measured if an excess in the semi-invisible decay of the Higgs to a photon and an invisible particle were to be discovered at the LHC. It was found that, if the dark photon was massless, an upper limit on its mass of a few GeV could be established in the best case scenario. For a $\text{BR}(h \to A A')$ of 1.5~\% and an integrated luminosity of 3~$\text{ab}^{-1}$, a $\chi^2$ upper limit on $m_{A'}$ of $\sim 8$~GeV and $\sim 22$~GeV could be found for gluon-fusion and $Z$-associated productions, respectively. If the dark photon was sufficiently massive, its mass could be determined potentially up to the sub-GeV level. For a $\text{BR}(h \to A A')$ of 1.5~\%, an integrated luminosity of 3~$\text{ab}^{-1}$ and $m_{A'}$ of 10~GeV, a $\chi^2$ uncertainty on $m_{A'}$ of $\sim $1.2~GeV and $\sim $5.1~GeV could be found for gluon-fusion and $Z$-associated productions, respectively. An alternative hypothesis that could explain the signal was that the Higgs decayed to a gravitino and a neutralino that then decayed to a photon and another gravitino. We found that this hypothesis could potentially be excluded at almost 95\% CL due to the effect of the Higgs width. Finally, we found that the presence of the signal $h\to AA'$ could easily be compatible with the absence of the $h\to Z A'$ signal.

We conclude with some statements about how our bounds could be improved in future work and other avenues worth studying. First, we did not consider any systematic errors. For $Z$-associated production, this is not expected to affect the results much, as the background is very small. However, the background is very large for the gluon-fusion channel. As such, we expect our bounds to be probably a bit optimistic, though an incorporation of the systematics in the test statistics as nuisance parameters should mitigate the difference~\cite{Cowan:2010js}. Second, there would certainly be ways to improve the background estimates. However, we found our results not very sensitive to the exact value of the background. For example, even in the extreme case that the background was twice as large, the contours of Fig.~\ref{fig:MaxMass} would only change by $\sim 20\%$. This is because these results come from comparing together different signals that are significant. Third, to maximize the reliability of our results, we were rather conservative in our cuts. More stringent cuts could potentially improve the results. For example, we could have cut on the missing transverse energy significance~\cite{CMS:2011bgj} or imposed a slightly stronger $p_T^{\text{miss}}$ cut. Fourth, we did not consider all Higgs production mechanisms. Namely, we did not consider the vector boson fusion production. Since the current limits from vector boson fusion are stronger than our projected $Z$-associated limits but weaker than our gluon-fusion limits, we expect vector boson fusion to give results in between those of these two channels. Finally, we mention that multiple Higgs production channels could in principle be combined to better constrain the properties of the dark photon.

\acknowledgments
This work was supported by the Ministry of Science and Technology of Taiwan under Grant No.~MOST-108-2112-M-002-005-MY3 and National Center for Theoretical Sciences, Taiwan.

\appendix

\section{Computation of the Higgs decay widths}\label{Sec:DecayWidths}

In this appendix, we compute the decay width of the Higgs to a dark photon and either a photon or a $Z$ boson when only kinetic mixing is present.

\subsection{Masses and mixing}\label{sSec:MassesAndMixing}

First, we discuss the Lagrangian and its diagonalization. We follow Ref.~\cite{Curtin:2014cca}, but simplify their results by assuming that the dark photon is lighter than the $Z$ boson. Consider the Lagrangian
\begin{equation}\label{eq:LagrangianMixing1}
  \mathcal{L}= -\frac{1}{4} W^3_{\mu\nu} W^{3\mu\nu} -\frac{1}{4}\hat{B}_{\mu\nu}\hat{B}^{\mu\nu} -\frac{1}{4}\hat{Z}_{D\mu\nu}\hat{Z}^{\mu\nu}_D + \frac{1}{2} \frac{\epsilon}{c_W}\hat{Z}_{D\mu\nu}\hat{B}^{\mu\nu} + \frac{1}{2}m_{D,0}^2 \hat{Z}_{D\mu}\hat{Z}_D^{\mu},
\end{equation}
where $ W^3_{\mu}$ is the neutral $SU(2)_L$ gauge field, $\hat{B}_\mu$ is the hypercharge gauge field, $\hat{Z}_{D\mu}$ the gauge field of a new $U(1)$ group and $\epsilon$ a mixing parameter. The mass $m_{D,0}$ could either come from the Stueckelberg mechanism or a dark Higgs with no difference as far as the phenomenology presented in this paper is concerned. The mixing term can be eliminated by a redefinition of the fields via
\begin{equation}\label{eq:Rotation1}
  \begin{pmatrix} W^3 \\ \hat{B} \\ \hat{Z}_D \end{pmatrix} = \begin{pmatrix} 1 & 0 & 0 \\ 0 & 1 & \frac{\epsilon}{c_W\sqrt{1- \frac{\epsilon^2}{c_W^2}}} \\ 0 & 0 & \frac{1}{\sqrt{1 - \frac{\epsilon^2}{c_W^2}}} \end{pmatrix} \begin{pmatrix} W^3 \\ B\\ Z_{D,0} \end{pmatrix}
  = R_1 \begin{pmatrix} W^3 \\ B \\ Z_{D,0} \end{pmatrix}.
\end{equation}
The mass Lagrangian in this basis is then 
\begin{equation}\label{eq:LagrangianMixing2}
  \mathcal{L}_M = \frac{1}{2} \begin{pmatrix} W^3_\mu & B_\mu & Z_{D,0,\mu}\end{pmatrix}
  \begin{pmatrix}
    \frac{g^2 v^2}{4} & -\frac{g g' v^2}{4} & -\frac{g g' \epsilon v^2}{4c_W\sqrt{1 - \frac{\epsilon^2}{c_W^2}}} \\
    -\frac{g g' v^2}{4} & \frac{{g'}^2 v^2}{4} & \frac{{g'}^2 \epsilon v^2}{4 c_W \sqrt{1 - \frac{\epsilon^2}{c_W^2}}}\\
    -\frac{g g' \epsilon v^2}{4c_W\sqrt{1 - \frac{\epsilon^2}{c_W^2}}} & \frac{{g'}^2 \epsilon v^2}{4 c_W \sqrt{1 - \frac{\epsilon^2}{c_W^2}}} & \frac{m_{D,0}^2 + \frac{{g'}^2 \epsilon^2 v^2}{4 c_W^2}}{1 - \frac{\epsilon^2}{c_W^2}}
  \end{pmatrix}
  \begin{pmatrix} W^{3\mu} \\ B^\mu \\ Z_{D,0}^\mu\end{pmatrix},
\end{equation}
where $g'$ ($g$) is the weak hypercharge ($SU(2)$) coupling and $v\approx 246$~GeV is the vacuum expectation value of the Higgs field. The mass matrix can be diagonalized by redefining
\begin{equation}\label{eq:Rotation2}
  \begin{pmatrix} W^3 \\ B \\ Z_{D,0} \end{pmatrix} = \begin{pmatrix} c_W c_\alpha & c_W s_\alpha & s_W \\ -s_W c_\alpha & -s_W s_\alpha & c_W \\ -s_\alpha & c_\alpha & 0 \end{pmatrix}\begin{pmatrix} Z \\ A' \\ A \end{pmatrix}
  = R_2 \begin{pmatrix} Z \\ A' \\ A \end{pmatrix},
\end{equation}
where $c_\alpha$ ($s_\alpha$) is the cosine (sine) of the angle $\alpha$, which is defined via
\begin{equation}\label{eq:alphaDefinition}
  \tan\alpha = \frac{-1 + \eta^2 s_W^2 + \delta^2 + \sqrt{(1 + \eta^2 s_W^2 + \delta^2)^2 -4 \delta^2}}{2\eta s_W},
\end{equation}
with
\begin{equation}\label{eq:etadeltaDefinition}
  \eta = \frac{\epsilon}{c_W\sqrt{1 - \frac{\epsilon^2}{c_W^2}}}, \qquad \delta = \frac{m_{D,0}}{m_{Z,0}\sqrt{1 - \frac{\epsilon^2}{c_W^2}}}, \qquad m_{Z,0}^2 = \frac{(g^2 + {g'}^2)v^2}{4}.
\end{equation}
The mass of the $Z$ boson and dark photon are
\begin{equation}\label{eq:mZmD}
  m^2_{Z, A'} = \frac{1 + \eta^2 s_W^2 + \delta^2 \pm \sqrt{(1 + \eta^2 s_W^2 + \delta^2)^2 - 4\delta^2}}{2} m_{Z,0}^2.
\end{equation}
Finally, we define for convenience
\begin{equation}\label{eq:Rdefinition}
  R = R_1 R_2.
\end{equation}

\subsection{$h \to AA'$}\label{sSec:HgammagammaD}

The amplitude takes the form\footnote{The computation was performed using \texttt{Package-X} \cite{Patel:2015tea}.}
\begin{equation}\label{eq:AmplitudeGeneral}
  M_{h \to AA'} = S^A(p_{A}\cdot p_{A'}g_{\mu\nu} - p_{A\mu}p_{A'\nu})\epsilon_A^{*\nu}\epsilon_{A'}^{*\mu},
\end{equation}
where the coefficient $S^A$ is given at one loop by
\begin{equation}\label{eq:SDefinition}
  S^A = S^A_W + \sum_f S^A_f.
\end{equation}
Consider a fermion $f$ of charge $Q_f$, mass $m_f$ and number of colours $N_C$. Label the weak hypercharge of its right-handed (left-handed) part as $Y_R^f$ ($Y_L^f$) and $T_3^f = Q_f - Y_L^f$. The fermion contribution to $h \to AA'$ is given at one loop by
\begin{equation}\label{eq:Sfgamma}
  S^A_f = -\frac{N_C Q_f(A_R^f + A_L^f) e g m_f^2}{8\pi^2(m_h^2 - m_{A'}^2)^2 m_W} F_1(m_h, m_{A'}, m_f),
\end{equation}
with
\begin{equation}\label{eq:ARALDefinition}
  A_R^f = g'Y_R^f R_{22}, \qquad A_L^f = g' Y_L^f R_{22} + g T_3^f R_{12},
\end{equation}
where $R_{ij}$ is the $ij$ element of the matrix $R$ and the function $F_1$ is defined as
\begin{equation}\label{eq:F1Definition}
  \begin{aligned}
  F_1(m_1, m_2, m_3) =\; & \left(m_1^2 - m_2^2\right)\left[ 2 + (-m_1^2 + m_2^2 + 4m_3^2)C_0(0, m_1^2, m_2^2;m_3,m_3,m_3)\right]\\
                         & + 2m_2^2\left(\Lambda(m_1^2;m_3,m_3) - \Lambda(m_2^2;m_3,m_3)\right).
  \end{aligned}
\end{equation}
The function $\Lambda$ is defined via
\begin{equation}\label{eq:LambdaDefinition}
  B_0(s;m_0,m_1) = \frac{1}{\epsilon'} + \ln\left(\frac{\mu^2}{m_1^2}\right) - \frac{1}{2s}\left(m_0^2 - m_1^2 + s\right)\ln\left(\frac{m_0^2}{m_1^2}\right) + \Lambda(s;m_0,m_1) + 2,
\end{equation}
where $B_0(s;m_0,m_1)$ is the scalar Passarino-Veltman function expressed using dimensional regularization in $d = 4- 2\epsilon'$ dimensions, and $C_0(s_1, s_{12}, s_2; m_0, m_1, m_2)$ the scalar three-point Passarino-Veltman function \cite{Passarino:1978jh}. The gauge boson loops contribute
\begin{equation}\label{eq:SWgamma}
  S^A_W = \frac{e g^2 R_{12}}{16\pi^2(m_h^2 - m_{A'}^2)^2 m_W^3} F_2(m_h, m_{A'}, m_W),
\end{equation}
where
\begin{equation}\label{eq:F2Definition}
  \begin{aligned}
     &F_2(m_1, m_2, m_3) = \\
         &\;\, \left(m_1^2 m_2^2 - 2m_1^2 m_3^2 + 2m_2^2 m_3^2 -12 m_3^4\right)
         \left(m_2^2 \Lambda(m_2^2; m_3, m_3) - m_2^2 \Lambda(m_1^2; m_3, m_3) - m_1^2 + m_2^2\right)\\
         &\;\, + 2m_3^2\left(m_1^2 - m_2^2\right)\left(m_1^2 m_2^2 - 6m_1^2 m_3^2 - 2m_2^4 + 6 m_2^2 m_3^2 + 12 m_3^4\right)C_0(0,m_1^2,m_2^2;m_3,m_3,m_3).
  \end{aligned}
\end{equation}
The decay width is then
\begin{equation}\label{eq:DecayWidthphoton}
  \Gamma^h_{AA'} = \frac{|S^A|^2(m_h^2 - m_{A'}^2)^3}{32\pi m_h^3}.
\end{equation}

\subsection{$h \to Z A'$}\label{sSec:HZgammaD}

If the dark photon is massive and the process is kinematically allowed, the Higgs will be able to decay to a dark photon and a $Z$ boson via a tree-level diagram of amplitude
\begin{equation}\label{eq:TreeLevelAmplitude}
  M_{\text{tree}} = \frac{v}{2}(g'R_{22} - g R_{12})(g'R_{21} - g R_{11})g_{\mu\nu}\epsilon_Z^{*\nu}\epsilon_{A'}^{*\mu} = \Omega g_{\mu\nu}\epsilon_Z^{*\nu}\epsilon_{A'}^{*\mu}.
\end{equation}
The coefficient $\Omega$ is however zero when the dark photon is massless, as expected from gauge invariance. As such, the loop contributions are important when the dark photon is very light and need to be considered. Since their analytical expressions are very complicated, we only present their value in the limit that the dark photon is massless. We have verified that the effect of the dark photon mass only impacts significantly the loop-level amplitude when it is negligible compared to the tree-level amplitude.

In the massless dark photon limit, the loop-level amplitude takes an analogous form to Eq.~\eqref{eq:AmplitudeGeneral} with $S^Z = S^Z_W + \sum_f S^Z_f$. The fermion loop contribution is
\begin{equation}\label{eq:SfZ}
  S^Z_f = -\frac{N_C Q_f \left(\hat{A}_R^f + \hat{A}_L^f\right) g^2 R_{12} m_f^2}{8\pi^2(m_h^2 - m_Z^2)^2 m_W} F_1(m_h, m_Z, m_f),
\end{equation}
where
\begin{equation}\label{eq:ARhatALhatDefinition}
  \hat{A}_R^f = g'Y_R^f R_{21}, \qquad \hat{A}_L^f = g' Y_L^f R_{21} + g T_3^f R_{11}.
\end{equation}
The gauge boson loop contribution is
\begin{equation}\label{eq:SWZ}
  S^Z_W = \frac{g^3 R_{11}R_{12}}{16\pi^2(m_h^2 - m_Z^2)^2 m_W^3} F_2(m_h, m_Z, m_W).
\end{equation}
The decay width is then
\begin{equation}\label{eq:DecayWidthZ}
  \Gamma^h_{ZA'} = \frac{\sqrt{(m_h^2 - (m_Z + m_{A'})^2)(m_h^2 - (m_Z - m_{A'})^2)}}{16\pi m_h^3}|\hat{M}|^2 ,
\end{equation}
where
\begin{equation}\label{eq:Mhat2Definition}
  \begin{aligned}
    |\hat{M}|^2 = &  \frac{|S^Z|^2}{2}\left[(m_h^2 - m_{A'}^2 - m_Z^2)^2 + 2m_{A'}^2 m_Z^2\right]
    + 3\text{Re}\{S^Z\}\left(m_h^2 - m_{A'}^2 - m_Z^2\right)\Omega \\
                  &+ \frac{\left((m_h^2 - m_{A'}^2 - m_Z^2)^2 + 8m_{A'}^2 m_Z^2\right)}{4m_{A'}^2 m_Z^2}\Omega^2.
  \end{aligned}
\end{equation}
For small $m_{A'}$, $\Omega \propto m_{A'}^2/v$ and the expression is well-behaved as $m_{A'}$ goes to zero.

\bibliography{biblio}
\bibliographystyle{utphys}

\end{document}